\documentclass[fleqn,usenatbib]{mnras}

\usepackage{newtxtext,newtxmath}

\usepackage[T1]{fontenc}

\DeclareRobustCommand{\VAN}[3]{#2}
\let\VANthebibliography\thebibliography
\def\thebibliography{\DeclareRobustCommand{\VAN}[3]{##3}\VANthebibliography}

\usepackage{graphicx}	
\usepackage{amsmath}	

\usepackage{subcaption}
\usepackage{array}
\usepackage{multirow}
\usepackage{booktabs}
\usepackage{xcolor}

\title[Scattering variations in pulsar timing]{Measuring scattering variations in pulsar timing observations: A test of the fidelity of current methods }

\author[A. D. Kulkarni et al.]{
A. D. Kulkarni,$^{1,2}$\thanks{E-mail: adkulkarni@swin.edu.au}
R. M. Shannon,$^{1,2}$
D. J. Reardon,$^{1,2}$ 
M. T. Miles$^{2,3}$
\\
$^{1}$Centre for Astrophysics and Supercomputing, Swinburne University of Technology, PO Box 218, Hawthorn, VIC 3122, Australia\\
$^{2}$ARC Centre of Excellence for Gravitational Wave Discovery (OzGrav), Mail H29, Swinburne University of Technology, PO Box
218,\\ Hawthorn, VIC 3122, Australia\\
$^{3}$Department of Physics and Astronomy, Vanderbilt University, 2301 Vanderbilt Place, Nashville, TN 37235, USA
}

\date{Accepted XXX. Received YYY; in original form ZZZ}

\pubyear{2024}

\begin{document}
\label{firstpage}
\pagerange{\pageref{firstpage}--\pageref{lastpage}}
\maketitle

\begin{abstract}
The turbulent nature of the ionised interstellar medium (IISM) causes dispersion measure (DM) and scattering variations in pulsar timing measurements. To improve precision of gravitational wave measurements, pulsar timing array (PTA) collaborations have begun the use of sophisticated and intricate noise modelling techniques such as modelling stochastic variations induced by the turbulent IISM and quasi-deterministic processes attributed to discrete structures. 
However, the reliability of these techniques has not been studied in detail, and it is unclear whether the recovered processes are physical or if they are impacted by misspecification. 
In this work, we present an analysis to test the efficacy of IISM noise models based on the data from the MeerKAT Pulsar Timing Array (MPTA) 4.5-year data release. We first performed multi-frequency, long-length (500 refractive length scale) simulations of multipath propagation in the IISM to study the properties of scattering variations under a variety of scattering conditions. The results of our simulations show the possibility of significant radio-frequency decorrelation in the scattering variations, particularly for the anisotropic scattering medium. 
Our analysis of the observed DM and scattering variations using the MPTA 4.5-year data set shows that there can be apparent anticorrelations between DM and scattering variations, which we attribute to the model fitting methods. 
We also report a possibility that plasma underdensities might exist along the sight lines of PSR J1431$-$5740 and PSR J1802$-$2124. 
Finally, using simulations, we show that the IISM noise models can result in the apparent measurement of strong frequency dependence of scattering variations observed in the MPTA data set.
Our analysis shows that improvements in the IISM noise modelling techniques are necessary to accurately measure the IISM properties.

\end{abstract}




\section{Introduction}
\label{Sec:Intro}

Pulsar timing arrays (PTAs) are Galactic-scale gravitational wave (GW) detectors that are sensitive to nanohertz-frequency GWs. The primary source of GWs at these frequencies is thought to be from the inspiral phase of supermassive black hole binaries \citep{1979ApJ...234.1100D,1990ApJ...361..300F}. The first GW detection at nanohertz frequencies is expected to come from a stochastic gravitational wave background (GWB), which produces a statistically common signal in the pulsar timing residuals among all pulsars along with a spatially correlated component \citep{1983ApJ...265L..39H}. PTA collaborations have found evidence for the presence of a common red noise signal that is statistically identical in all pulsars \citep{2020ApJ...905L..34A,2021ApJ...917L..19G,2021MNRAS.508.4970C,2022MNRAS.510.4873A,2023RAA....23g5024X}. They have also reported evidence of angular correlations that are consistence with a GWB \citep{2023A&A...678A..50E,2023ApJ...951L...6R,2023ApJ...951L...8A,2023RAA....23g5024X,2025MNRAS.536.1489M}. However, apart from the GWB, there are other sources of noise that are routinely observed in the PTA data sets. An accurate and complete knowledge of these noise signals is critical, as the detection statistic of the GWB is sensitive to the strengths of noise signals in PTA data sets \citep{2009PhRvD..79h4030A}.

In the timing of millisecond period pulsars (MSPs) at radio wavelengths, the most dominant source of noise is usually due to the interaction of the radio waves with the intervening ionised interstellar medium (IISM) \citep{1990ApJ...364..123F,2010arXiv1010.3785C}. 
Radio waves experience a frequency-dependent dispersion when propagating through free electrons in the ionised interstellar medium.
Additionally, the radio waves are scattered off of the inhomogeneities in the electron density. These interactions can result in radio waves travelling along an indirect line of sight propagating to the observer, which cause delays in the arrival times of the pulses at the telescope. Because of the motion of the line of sight through the turbulent IISM, these delays vary stochastically and can affect the sensitivity of a PTA to GWs. Both dispersion and scattering have a high degree of chromaticity (dispersion delays are proportional to $\nu^{-2}$  and scattering delays are thought to be proportional to $\approx \nu^{-4}$)
and hence can be measured and removed with the use of multi-frequency measurements of the times of arrival. The net delay in the arrival times consists of a variety of terms originating from these IISM interactions \citep{2010arXiv1010.3785C,2018CQGra..35m3001V}. The strongest of all is due to the variation in the dispersion measure (DM), which is the total electron column density along the line of sight ($ \mathrm{DM} = \int_0^L n_{e} dl$, where $n_e$ is the electron density and $L$ is the distance to the pulsar). The other delay terms are collectively referred to as `scattering noise' and they include long-term delays due to multipath propagation and angle of arrival variations amongst others \citep{2010arXiv1010.3785C}. Scattering noise contributions are often subdominant for many pulsars and hence PTA collaborations have been treating DM variation as the primary source of IISM noise \citep[and sometimes as the only noise term][]{2016MNRAS.457.4421C,2016MNRAS.458.1267V}. However, with the use of multi-frequency or wide-band observations and more sensitive telescopes and receiving systems, it is possible to detect the subdominant scattering variations. 
For example, the observations of extreme scattering events, which are the unusual variations in the brightness of the pulsars, have shown that accounting for scattering delays can become important for high precision timing experiments \citep{2015ApJ...808..113C,2018MNRAS.474.4637K}.  

In addition to causing dispersion and scattering delays, propagation through the IISM can also be responsible for the scintillation of pulsars. The interference pattern of the scattered wavefront coupled with motion of line of sight is seen as flux density variations both across frequency and time. The resulting dynamic spectrum of the flux variation shows islands of correlated flux, known as `scintles', which can be characterised by the scintillation time scale $\Delta t_{d}$ and scintillation bandwidth $\Delta \nu_{d}$. The temporal variations in these quantities can be used to study the properties of the IISM. For example, combined measurements of the scintillation timescale and DM variations from long-term timing of pulsars have long been used to constrain the power law spectrum of turbulence, $P_{\delta n_e} (q) \propto q^\zeta $ \citep{2007MNRAS.378..493Y,2013MNRAS.429.2161K,2017ApJ...841..125J} where $q$ is the electron-density wavenumber. Such observations indicate that the IISM can often be characterized by a Kolmogorov spectrum of turbulence with a spectral steepness of $\zeta=-11/3$ \citep{1995ApJ...443..209A,2013MNRAS.429.2161K}. However, deviation from the Kolmogorov spectrum have also been noted in many cases \citep{1994ApJ...430..581F,2007ASPC..365..299W}. The measurements of the frequency dependence of scintillation properties such as the scintillation bandwidth can give additional insights into the nature of turbulence \citep{2017MNRAS.470.2659G,2021ApJ...917...10T,2022A&A...664A.116L}.
The temporal flux density variations of the pulsar are useful to study the turbulent induced plasma structures in the IISM \citep{2021MNRAS.501.4490K}. Apart from improving the significance and robustness of a GW detection, the long-term modelling of IISM delay terms can itself provide a unique opportunity to study the structure of the IISM on a wide range of spatial scales \citep[$10^9 - 10^{17}$ m,][]{1995ApJ...443..209A,2016Sci...351..354B}. As PTA collaborations observe pulsars over many years with relatively high $\sim$ week-month cadences, these observations can be used to capture structures in the IISM at different length scales.
The measurements of DM variations and scattering variations from PTA data sets can open up new avenues for understanding the physics behind the morphology of the IISM.

Currently, when modelling IISM contributions, a common approach is to treat DM and scattering variations as independent Gaussian processes that follow a power law in the Fourier domain \citep{2013PhRvD..87j4021L,2024ApJ...972...49L}. The DM noise term has a fixed radio frequency scaling of $\nu^{-2}$. While in many cases the chromaticity of scattering noise is fixed to $ \nu^{-4}$  \cite[e.g.,][]{2023ApJ...951L...6R}, it is also occasionally  treated as a free parameter, $\beta$ \cite[][]{2025MNRAS.536.1467M}. However complex chromatic features in the timing residuals have motivated the use of other deterministic signals. For example, recently the Parkes Pulsar Timing Array (PPTA) and the MeerKAT Pulsar Timing Array (MPTA) have included additional signals such as chromatic Gaussian events or annual chromatic variations to properly account for discrete structures in the IISM \citep{2021MNRAS.502..478G,2023ApJ...951L...7R,2025MNRAS.536.1467M}.
The presence of discrete structures can potentially be attributed to extreme scattering events or other plasma intermittencies \cite[][]{1987Natur.326..675F,2015ApJ...808..113C}. Another important chromatic signal arises from electron column density variations induced by the solar wind, which is thought to be especially important for pulsars close to the ecliptic plane \citep{2022ApJ...929...39H,2024A&A...692A..18S}. As the complexity of the noise models increases, PTA collaborations face new challenges. It is often the case that the noise processes present in pulsar timing data are covariant. Model misspecification can potentially cause inference bias \citep{2023ApJ...956...14D,2024MNRAS.532.4026D}. In their recent analysis, the MPTA reported presence of unusual chromatic noise in some of its pulsars. The chromaticity of scattering variations was found to be as high as $8.8^{+1.9}_{-1.1}$ \citep{2025MNRAS.536.1467M}, whereas the steepest expected chromatic index has been predicted to be only $\approx 6.4$ \citep{2017MNRAS.464.2075S}.

The goal of this work is to address the challenges outlined above and to assess the performance of currently used IISM noise modelling techniques.
We first perform simulations of the propagation of radio waves in the IISM. In particular, we explore the effect of anisotropy on the temporal and spectral correlation of scattering delays. Section \ref{Sec:Simulations} reports the results of our simulations, including predictions for scattering delays from propagation in a Kolmogorov medium. Here, we also explore the relation between DM variations and scattering variations. These results form a foundation for our analysis on the chromatic noise variations in the MPTA 4.5-year data set, which we present in section \ref{Sec:MPTA_Data}. We conclude in sectioin \ref{Sec:Chromaticity} by investigating potential causes for the unusually steep chromaticity of the scattering noise signals.

\section{Background}
\label{Sec:Background}
Previous studies of pulsars in the Milky Way have highlighted the complexity and diversity in the structures of the IISM along different lines of sight \citep{1987Natur.326..675F,1994MNRAS.269.1035G,1995ApJ...443..209A,2001ApJ...549L..97S,2016MNRAS.462.2587G,2024MNRAS.528.6292J,2024MNRAS.527.7568O}. However, the most widely applied model to explain the scintillation assumes the scattering of radio waves through thin screens \citep{1986ApJ...310..737C,2001ApJ...549L..97S,2006ApJ...637..346C,2016ApJ...817...16C}. In this approach, plasma causing scattering effects is assumed to be concentrated in one or more two-dimensional \cite[and in some cases one-dimensional,][]{2024MNRAS.528.6292J} screens of negligible thicknesses, located between a compact object such as a pulsar and the Earth.
The electron density in such screens can have variations at different spatial scales, due to which radio waves propagating through the screen experience position dependent phase variations $\phi(r)$. These are best represented by a structure function defined as  
\begin{equation}
    D_{\phi}(\Delta r) = \,<(\phi(r) - \phi(r+\Delta r))^2>,
    \label{eq:phase_structure_function}
\end{equation}
where $<\dots>$ represents ensemble average and $\Delta r$ is the displacement on the screen \citep{1992RSPTA.341..151N}. This representation is useful for a variety of stationary random processes used to describe the phase variations. However, the most attractive and commonly used representation of phase variations is a power-law  of their magnitudes in the spatial frequency domain specified by a wavenumber $q$. 
The power spectrum of the phase variations is normally modeled using the form \citep{1986ApJ...310..737C,1990ARA&A..28..561R,1999ApJ...517..299L}
\begin{equation}
    P_\phi (q) = C_{\phi}^{2}q^{-\zeta}
    \label{eq:plasma_powerlaw}
\end{equation}
in the range $q_{\rm out} \leq q \leq q_{\rm in}$, where $ q_{\rm out}$ and $q_{\rm in}$ are, respectively, the outer-scale and the inner-scale of the turbulence, $C_{\phi}$ is the strength in phase fluctuations and $ 3 > \zeta > 4$. For such a medium, the phase structure function is also a power-law of the form $D_{\phi}(\Delta r) = (r/s_d)^{-\zeta +2}$. Two scales from the power spectrum become important in the study of scattering. The first is the diffractive scale $s_{d}$ at which the phase variations in the screen are one radian, such that $D_{\phi}(s_d) = 1$. Plasma inhomogeneities comparable to this scale cause diffraction of the radio waves. The second scale is the refractive scale which is related to $s_{d}$ through $s_{\rm ref} = r_{f}^2/s_{d}$ where $r_{f} = \sqrt{ z/k}$ is the Fresnel scale for a screen at a distance of $z$ and $ k = 2\pi/\lambda$ where $\lambda$ is the wavelength. Plasma variations comparable to or more than $s_{\rm ref}$ are responsible for the refraction of radio waves. Both of these effects produce measurable changes in the properties of a pulsar such as its flux density \citep{2023MNRAS.526.3370G}, DM \citep{2007MNRAS.378..493Y,2017ApJ...841..125J}, rotation measure \citep{1990ApJ...363..515L}, and morphology of the pulse profile \citep{2008ApJ...674L..37H}. Temporal variations in these quantities can be used to constrain the exponent of the power-law in equation \ref{eq:plasma_powerlaw}. Evidence from earlier studies suggests a power spectrum exponent close to the Kolmogorov value of $ \zeta=11/3$ \citep{1995ApJ...443..209A}. However deviations from the Kolmogorov spectrum at larger spatial scales have been observed \citep{1994ApJ...430..581F,2007ASPC..365..299W}.

Temporal variations in propagation effects are a result of the motion of effective line of sight of the pulsar through the turbulent plasma, as the pulsar, Earth, and IISM move. For example, the motion through the interference pattern of the scattered wavefront at the observer's plane is seen as flux density variations. 
On short time scales (usually smaller than a few hours for pulsars observed at decimetre wavelengths), diffractive effects dominate and causes modulations in the flux dynamic spectrum. In this regime, scattering causes multipath propagation of the radio waves, as a result of which the observer receives rays from multiple directions. The angular size of the image of the source is broadened and can be given by $\theta_{d} \approx 1/ks_{d}$. The multipath propagation is equally responsible for scatter broadening of the pulse shape in time domain \citep{1969Natur.221...31S}. 
The scatter broadening distorts the pulse profile as the radiation scattered from larger angles has a longer path than the direct line of sight. This results in a shift in the centroid of the resultant pulse and introduces an extra delay in the arrival time of the pulse where the mean is equal to the scattering time scale $\tau_{s} = \theta_{d}^2z/c$ \citep{2008ApJ...674L..37H}, where $c$ is the speed of light. For moderately scattered pulsars, $\tau_{s}$ is less than the emitted pulse profile width and potentially the resolution of the recording instrument. This effect is more pronounced in the case of strongly scattered pulsars, especially when observed at lower radio frequencies \citep{2010arXiv1010.3785C}.
Some of the previous studies have attempted to estimate these variations by measuring the scattering delays at each epoch \cite[e.g.][]{2021ApJ...917...10T,2022A&A...664A.116L,2023MNRAS.525.1079M}. 

On longer time scales of weeks to months, refractive effects dominate. Structures larger than the refractive scales $s_{\rm ref}$ act like lenses and focus or de-focus the radiation. This is observed as slow temporal variations in the flux density of the pulsar. Another result of the refraction is the displacement of the location of the image center from the direct line of sight \citep{1986ApJ...310..737C}. This would result in an apparent angular shift in the position of the pulsar by an amount equal to the refractive angle $\theta_{r}$, which is related to the gradient of the phase as $\theta_{r} = \nabla \phi/k$. The angular shift in the image position implies that the pulses experience an extra time delay relative to the direct line of sight,  \begin{equation}
    \centering
    \Delta t_{r} \sim \frac{\theta_{r}^2 z}{2c}.
    \label{eq:t_ref}
\end{equation}
This delay is the consequence of the path length difference between the direct line of slight and the centre of the refracted image.  For a Kolmogorov medium, $\theta_{r} < \theta_{d}$ on average; however, the presence of any large-scale structures deviating from the Kolmogorov fluctuations would increase the size of the refractive delay.
 
For PTA observations, the largest radio-frequency dependent contribution to arrival time variations is due to the DM variations \citep{2007MNRAS.378..493Y,2017ApJ...841..125J}, which has a radio frequency dependence of $\nu^{-2}$. However, the radio frequency dependence (chromaticity) of the delay terms due to scattering depends on the characteristics of the turbulence in the IISM. For a Kolmogorov medium ($\zeta = 11/3$), $\tau_{s}$ and $\Delta t_{r}$ terms have chromaticity of $\nu^{-4.4}$ \citep{1977ARA&A..15..479R}. Although these two effects have gained significant attention, other terms such as the image averaged DM delay or changes in the scatter broadening time due to refractive interstellar scintillation have not been studied in as much detail. An extensive discussion on the strength and the radio frequency dependence of various delay terms due to the IISM can be found in \citet{2010ApJ...717.1206C}, \citet{2010arXiv1010.3785C}, and \citet{2017MNRAS.464.2075S}. 

\section{Simulations}
\label{Sec:Simulations}

Simulations of multipath propagation have provided useful insights in understanding the structure of the IISM \citep{1986ApJ...310..737C,1990ApJ...364..123F,2010ApJ...717.1206C,2017MNRAS.464.2075S}. Large-scale structures are responsible for refractive effects, whereas smaller scale structures are responsible for diffractive effects. Such simulations are complicated by the large dynamic range of spatial scales required to fully model refractive and diffractive effects. Moreover, a larger dynamic range is necessary to study stronger scattering as the difference in spatial scales responsible for refractive and diffractive effects increases. In such cases, it has often been necessary to make comprises and focus on studying one effect.  
For modelling refractive effects, it is common to assume ensemble average properties for diffractive effects (such as the size of the scatter broadened image or pulse broadening time), which can often be calculated analytically \citep{1999ApJ...517..299L}.

Early simulations such as by \citet{1986ApJ...310..737C} modelled one-dimensional screens in which the total phase variation was treated as the combination of long-term refractive phase structures and the small-scale diffractive phase structures assuming ensemble average statistics. The results from those demonstrated the phenomenon of the refractive image wandering and its effect on the pulsar dynamic spectra. Similar one-dimensional screens were further used by \citet{1990ApJ...364..123F} to investigate the propagation effects in pulsar timing analysis.
Later studies extended the scattering analysis on full two-dimensional screens. 
\citet{2010ApJ...717.1206C} and \citet{2017MNRAS.464.2075S} both considered refractive effects and the impact of second-order refractive variations on diffractive effects. While \citet{2017MNRAS.464.2075S} focused on understanding refractive effects by assuming ensemble-averaged properties for diffractive effects, 
\citet{2010ApJ...717.1206C} presented complete simulations with both diffractive and refractive terms sampled. The complete simulation limits the size of the phase screens that can be modeled. The simulations of \citet{2010ApJ...717.1206C} were tailored to understand the IISM in the direction of the millisecond pulsar J0437$-$4715, which is one of the weakly scattered pulsars, thus requiring a lower dynamic range.

With the advancement in computing, it is now possible to generate fully diffractive simulations that show refractive effects over a wide range of strengths of scattering. Building on previous efforts, we have simulated large two-dimensional screens which capture many refractive scales ($500\,s_{\rm ref}$) along with smaller diffractive scales. First, we consider refractive effects arising naturally from the Kolmogorov spectrum. We investigate the role anisotropy plays in affecting propagation effects. We also simulate the presence of underdensities in the plasma screens by introducing a two-dimensional Gaussian structure in addition to the underlying Kolmogorov turbulence. 

\subsection{Simulations of refractive scattering} 
We use the python package \textsc{scintools} \citep{2023MNRAS.521.6392R} to simulate the scattering due to large-scale IISM structures.  The package builds on the code presented in \citep{2010ApJ...717.1206C}. It first generates a single thin phase screen based on parameters describing the characteristics of the turbulence, such as the exponent of the power law, strength of scattering, and properties of the anisotropy. Following the approach mentioned in \citet{2010ApJ...717.1206C}, the strength of scattering in our simulations is parametrized using the Born variance parameter $m_b^2$. This parameter provides a good approximation of the variance of the flux density for weak scintillation when $m_b^{2} < 1$. Although for strong scintillation, $m_b^2$ deviates from the actual variance of the flux density, it can still be a useful parameter to define the strength of scattering. For a Kolmogorov medium, the phase structure function at Fresnel scale $r_f$ is related as $m_b^2 = 0.773D_{\phi}(r_f)$ \citep{2010ApJ...717.1206C}. The scintillation becomes progressively stronger when $m_{b}^2 > 1$. The anisotropy for a two-dimensional screen can be defined by the axial ratio parameter $a_r$ for an ellipse and the orientation angle $ \psi$ of the major axis with respect to the $x$ axis. In our simulations, the $x$ axis is also the direction of transverse motion of the line of sight, whereas the perpendicular direction to the $x$ axis is the $y$ axis.  
As the refractive scale $s_{\rm ref}$ increases rapidly with $m_b^2$, simulating long screens of $\sim 500\, s_{\rm ref}$ while maintaining enough resolution to capture diffractive effects becomes computationally and memory intensive.
Thus we restricted our simulations to the scattering strength of $ m_b^{2} = 20$ with phase variations conforming to the Kolmogorov spectrum ($\zeta = 11/3$). Although the strength of scattering in our simulations may not be comparable to strongly scattered pulsars, the strong scattering regime still applies and the results can be extrapolated for higher scattering strengths.
Instead in this work, we focus on varying the degree and properties of anisotropy to assess its impact on scattering noise. The simulations were carried out using axial ratios of $ a_r = 1, 2, 5$ and orientation angles of $ \psi = 0^\circ, 45^\circ, 90^\circ, 135^\circ$. 
Motivated by the procedure followed by \citet{2010ApJ...717.1206C}, the dimensions of the screen and the grid spacing were chosen such that the simulation can capture both diffractive and refractive effects. To achieve this, the grid separation was set to $\Delta r = s_d/4$.    
The length of the screen along the direction perpendicular to the motion was set to $ L_{y}=10 s_{\rm ref}$. In the direction parallel to motion, the screen size was chosen to be $L_{x}=500 s_{\rm ref}$. 
The simulations assume that a parallel wavefront impinges on the phase screen and subsequently the electric field at the observation plane is calculated. We ran these simulations with the default settings of the fractional bandwidth of $0.25$ centred at $ \nu_{c} = 1400$\,MHz (corresponding to the bandwidth of $ \Delta \nu = 350$\,MHz ) available in \textsc{scintools} package. This choice of frequency matches the simulations carried out by \citet{2010ApJ...717.1206C}, which allows us to compare the results. It is also relevant because it coincides with PTA observations performed at the L-band

\subsection{Measurement of flux and centroid variations}
\label{subsec:measurement_of_scattering_variations}
A simulation produces a three-dimensional matrix of the resultant electric field $ E(x,y,\nu)$ in the plane of observation, at a radio frequency $\nu$. This information can be used to obtain the flux density variation using the relation 
\begin{equation}
    S_{\nu_{c}}(x,y) = \int_{\nu_{c} - \Delta \nu/2}^{\nu_{c} + \Delta \nu/2}|E(x,y,\nu)|^{2} d\nu.
    \label{eq:flux}
\end{equation}
Similarly, the electric field can also be used to construct a pulse broadening function (PBF) at a center frequency $\nu_{c}$ using the relation   
\begin{equation}
    I_{\nu_{c}}(x,y,t) = |\mathscr{F}(E(x,y,\nu))|^{2},
    \label{eq:impulse_from_efield}
\end{equation}
where $\mathscr{F}$ denotes the Fourier transform operation.

The IISM delays the time of arrival and broadens the shape of the pulse. Under conditions of strong scattering, such as those simulated in this work, the mean shift regime applies \citep{2010arXiv1010.3785C}. The arrival time of the pulse in the mean shift regime can be approximated by the perturbation caused in the centroid of the pulse \citep{2008ApJ...674L..37H}, which can be calculated as 
\begin{equation}
    \tau(x,y,\nu_{c}) = \frac{\int_{-\infty}^{\infty} t I_{\nu_{c}}(x,y,t) dt}{\int_{-\infty}^{\infty} I_{\nu_{c}}(x,y,t) dt}.
    \label{eq:centroid}
\end{equation}
We assume the motion of line sight through the middle of the screen along the $x$ direction. Thus, for the rest of this work, the centroids were measured at the position $y_c = L_{y}/2$.

The resultant DM contribution is expected to come from the centre of the scattered image of the source at each radio frequency. In the presence of refractive effects, the image center wanders away from the direct line of sight in a manner that is also radio-frequency dependent. Thus the dispersion delay is also expected to be frequency dependent as the location of image centres varies with frequency. Moreover, at each frequency the resultant dispersion delay would be due to the averaging over all possible ray paths from the scattering disc. A detailed discussion of this effect can be found in \citet{2016ApJ...817...16C}. Both of these effects are responsible for a frequency-dependent DM in the simulations. However, in this work our approach is to treat the DM as constant in radio frequency arising from  dispersion from the direct line of sight in the phase screen. Under this condition the DM variation in the simulation can be obtained from the direct line of sight phase $\phi$ using the relation 
\begin{equation}
    {\rm DM}(x) = -\frac{\nu }{c r_e} \phi(x,y=L_{y}/2),
    \label{eqn:DM_from_phases}
\end{equation}
where $r_e$ is the classical radius of an electron.
To get the variations in the subdominant scattering term, we need to correct for the dispersion sweep directly from the electric field prior to computing the PBFs. This is achieved by multiplying the electric field by a phase correcting filter based on the expected dispersion delay from the direct line of sight \citep{1975MComP..14...55H}. This DM-corrected (i.e., coherently dedispersed) electric field can be used to compute the PBFs according to equation \ref{eq:impulse_from_efield}. Here we note that the PBFs computed in this way may still retain a subdominant frequency-dependent residual dispersive delay along with the delay arsing from scattering effects. This may alter the chromaticity of the resultant signal. As the diffractive scale is well sampled in these simulations, the PBFs may have contributions from both refractive and diffractive effects. The diffractive effects cause fluctuations in instantaneous PBFs \citep{2010ApJ...717.1206C}, which are departures from the ensemble average PBF. This phenomenon is known as the finite scintle effect \cite[][]{2010arXiv1010.3785C}. To mitigate these, the instantaneous PBFs were averaged over one refractive scale $S_{\rm ref}$ and then used to compute the times of arrival along the $x$ axis across the simulation length using the relation \ref{eq:centroid}. The measurements obtained in this way are referred to as ``centroid variations'' and are expected to represent the scattering variations in the PTA data sets.

\subsection{Sub-banded centroid variations}
\label{subsec:Sub-banded scattering variations}
The availability of the multi-frequency observations of pulsars at the same epoch makes it possible to measure the chromaticity of the scattering noise variations in PTA observations. A similar strategy can be used in simulations to measure the radio frequency dependence of the scattering noise. To do this, we divide the simulated bandwidth of $350$\,MHz into $32$ frequency channels and compute a separate PBF at each channel following equation \ref{eq:impulse_from_efield}. Once again to mitigate fast variations, the PBFs were averaged over one refractive scale computed at the center frequency. After this the centroid values were computed in each frequency channel using the equation \ref{eq:centroid}. This method allows us to measure the chromaticity of the centroid variations in our simulations and to compute the correlation of the centroid variations between frequency channels for different conditions of anisotropy. 

\begin{figure*}
    \centering
    \includegraphics[width=\linewidth]{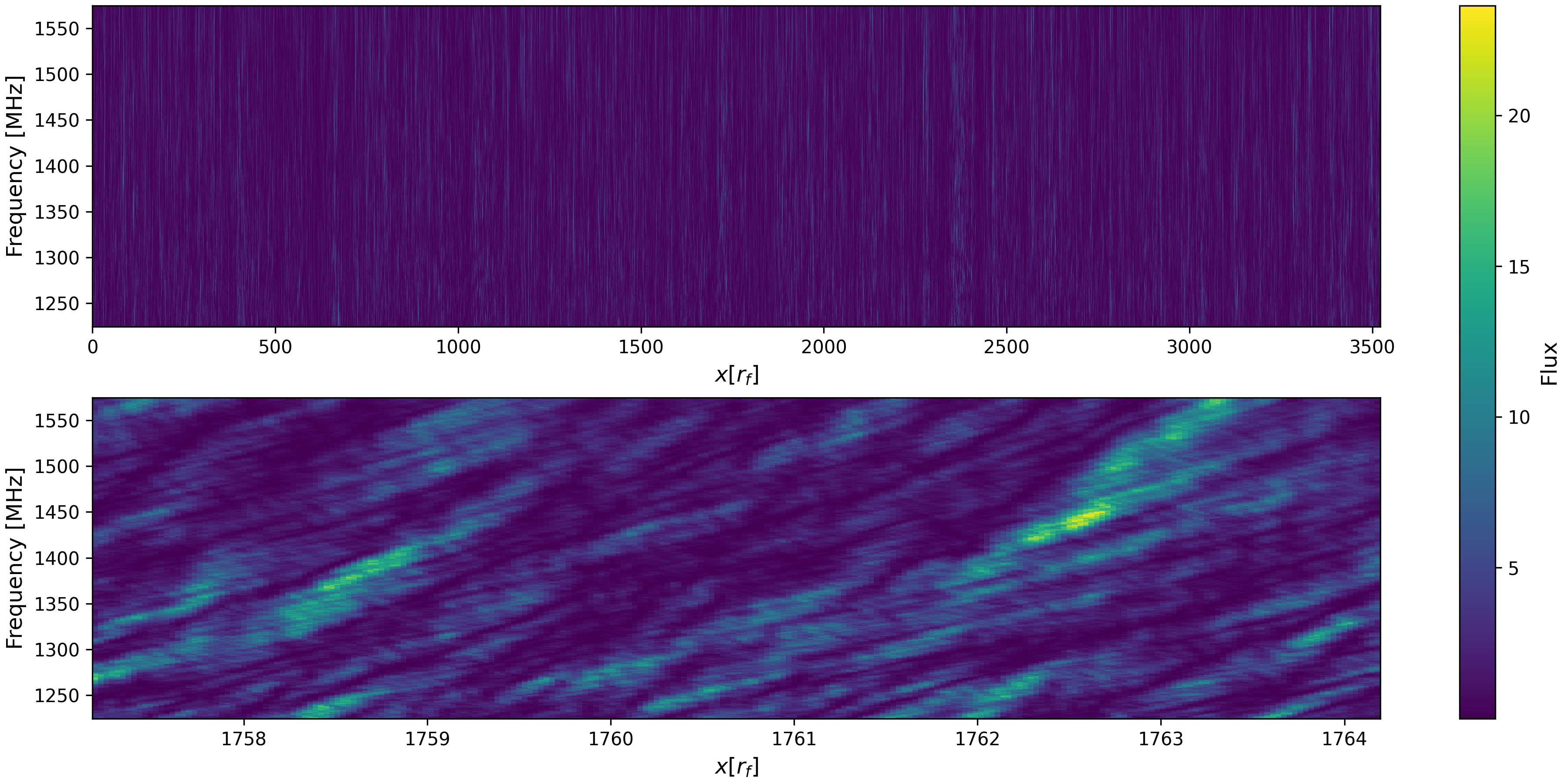}
    \caption{Dynamic spectra for an isotropic scattering screen. The top panel shows the dynamic spectrum for the entire simulation spanning a length of $ 500\,S_{\rm ref}$ in the $x$ direction. The bottom panel shows the zoom in of the top panel, spanning a length of $ 1\,S_{\rm ref}$. The tilting of the structure seen in the bottom panel is the result of the presence of refractive phase gradients in the screen. The right panel shows the flux density in arbitrary units as defined in equation \ref{eq:flux}.}
    \label{fig:Dynamic spectrum}
\end{figure*}

\begin{figure*}
    \begin{subfigure}[t]{.47\linewidth}
        \centering\includegraphics[width=1\linewidth]{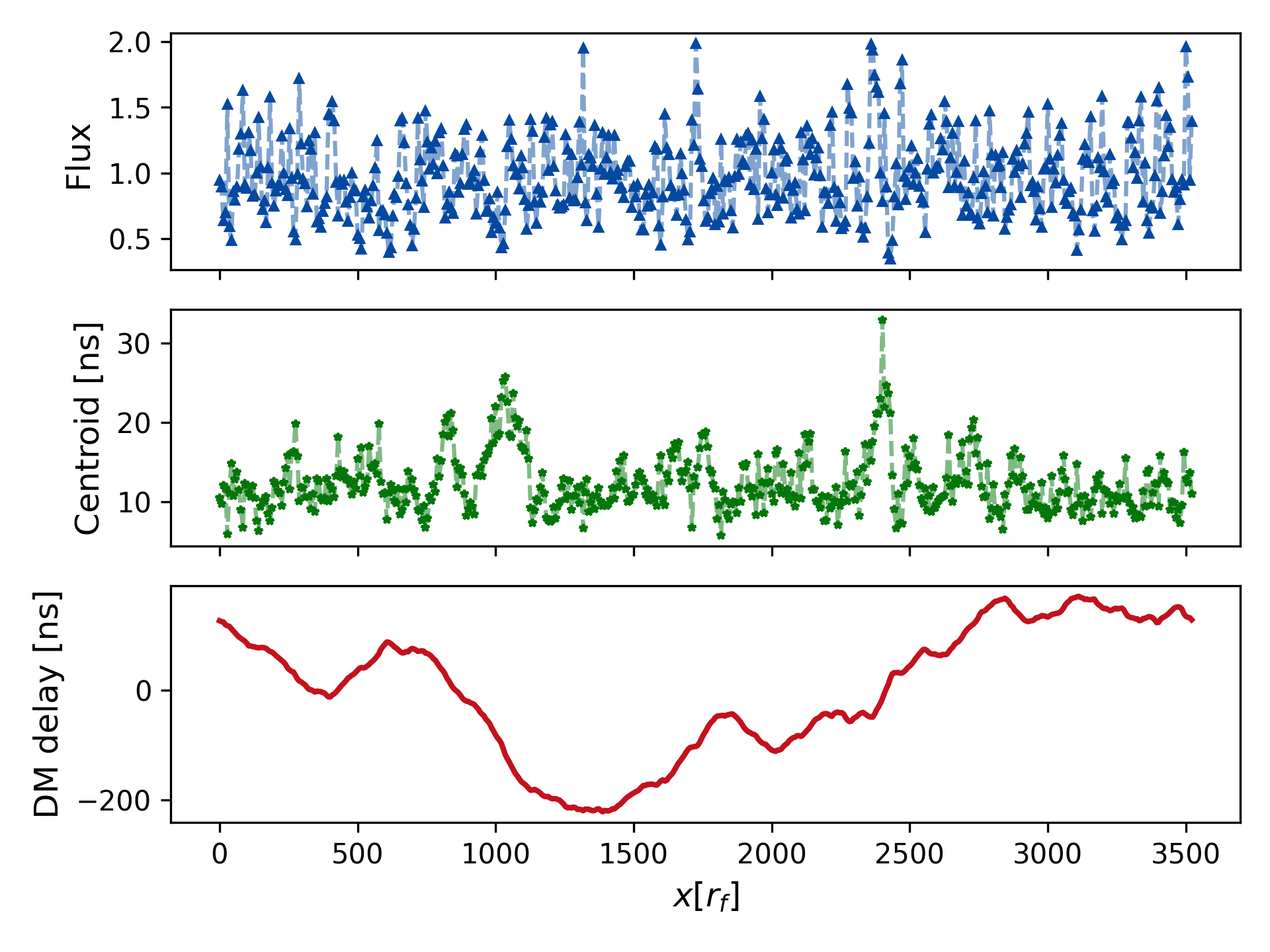} 
        \caption{}
        \label{subfig:iDM_centroid_flux}
    \end{subfigure}
    \begin{subfigure}[t]{.47\linewidth}
        \centering\includegraphics[width=1\linewidth]{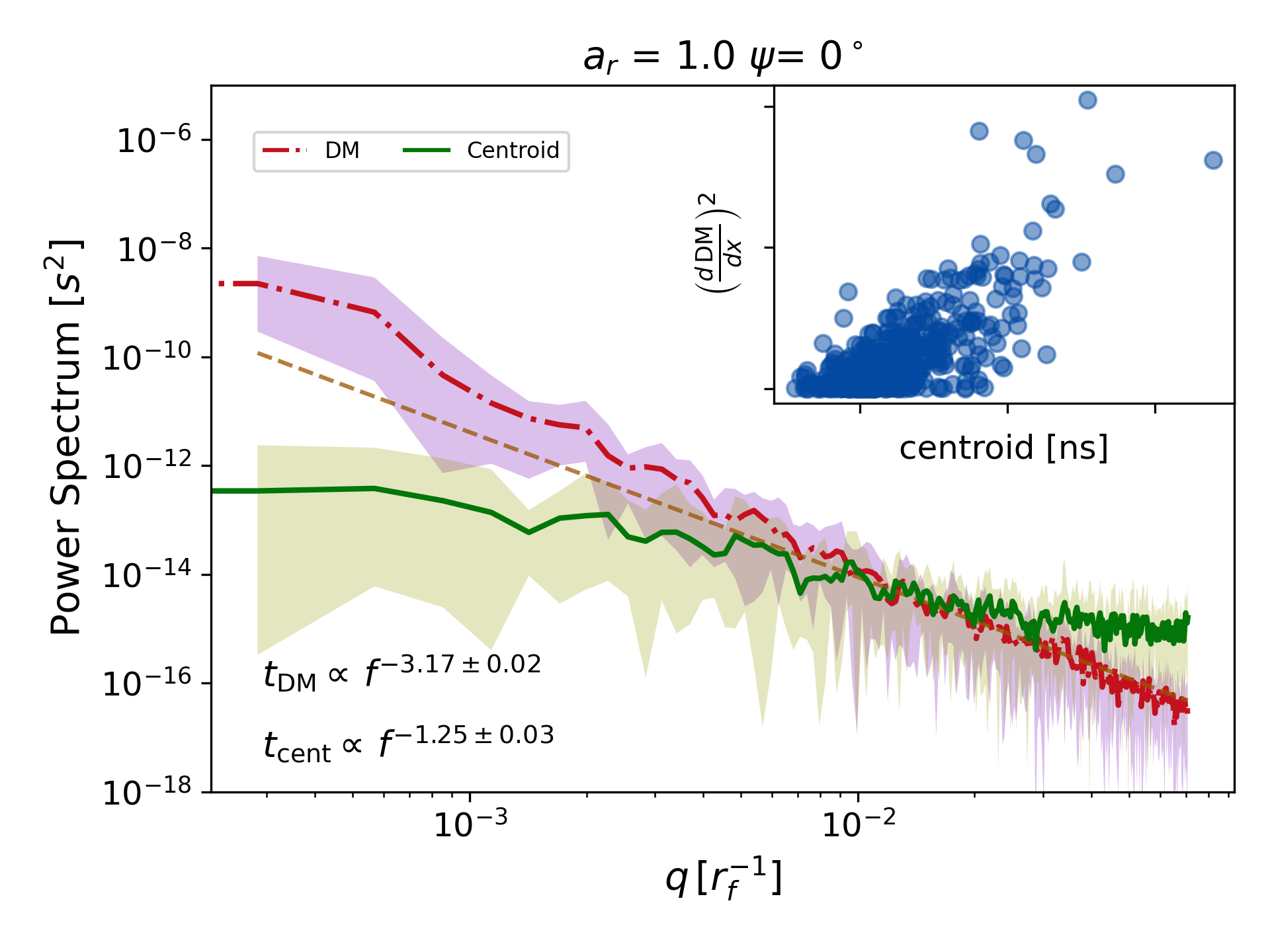} 
        \caption{}
        \label{subfig:corr_chrom_sqdmdr_simulation}
    \end{subfigure}
    \begin{subfigure}[t]{.47\linewidth}
        \centering\includegraphics[width=1\linewidth]{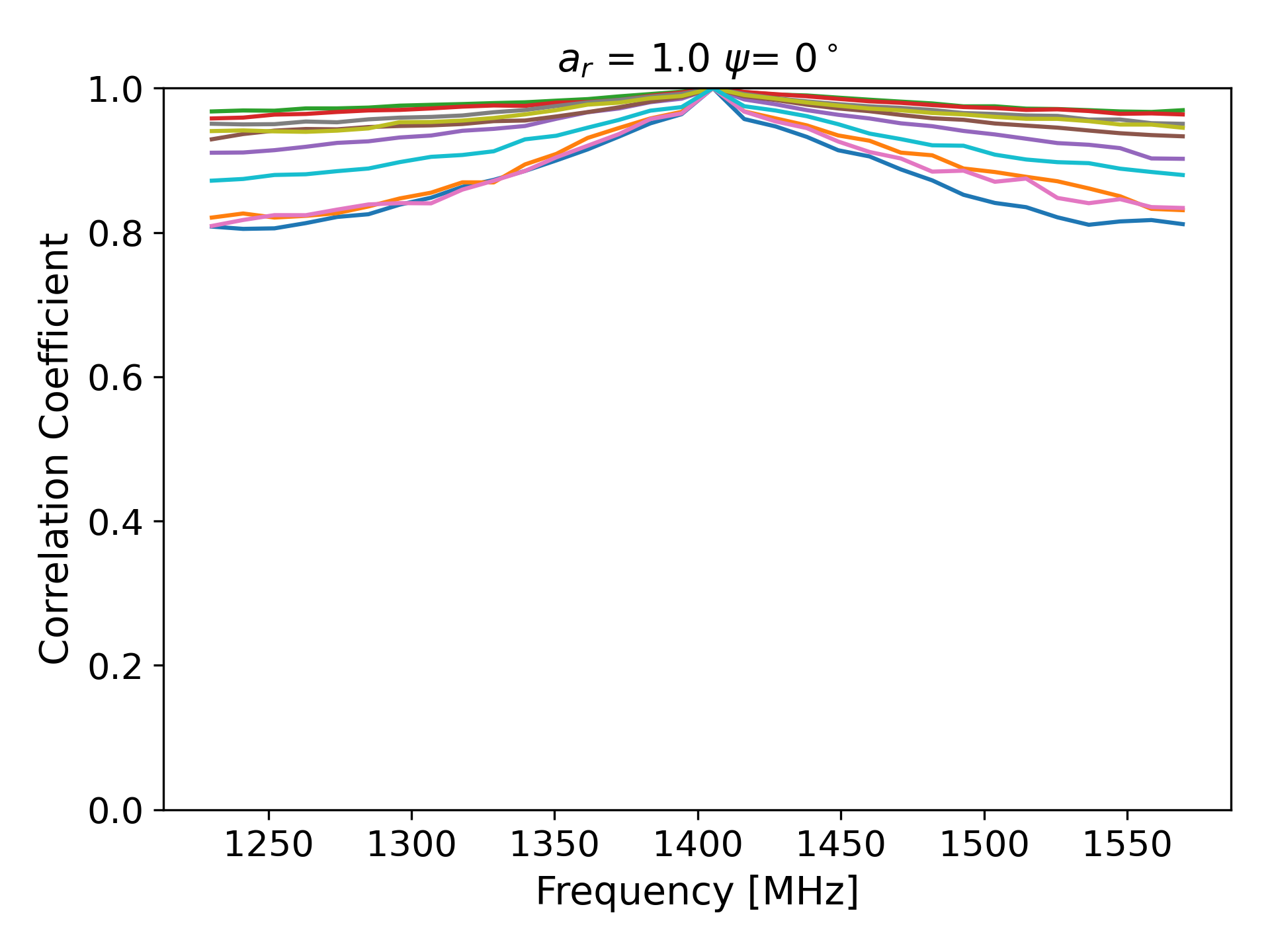}
        \caption{
        }
        \label{subfig:centroid_freq_corr_iso}
    \end{subfigure}
    \begin{subfigure}[t]{.47\linewidth}
        \centering\includegraphics[width=1\linewidth]{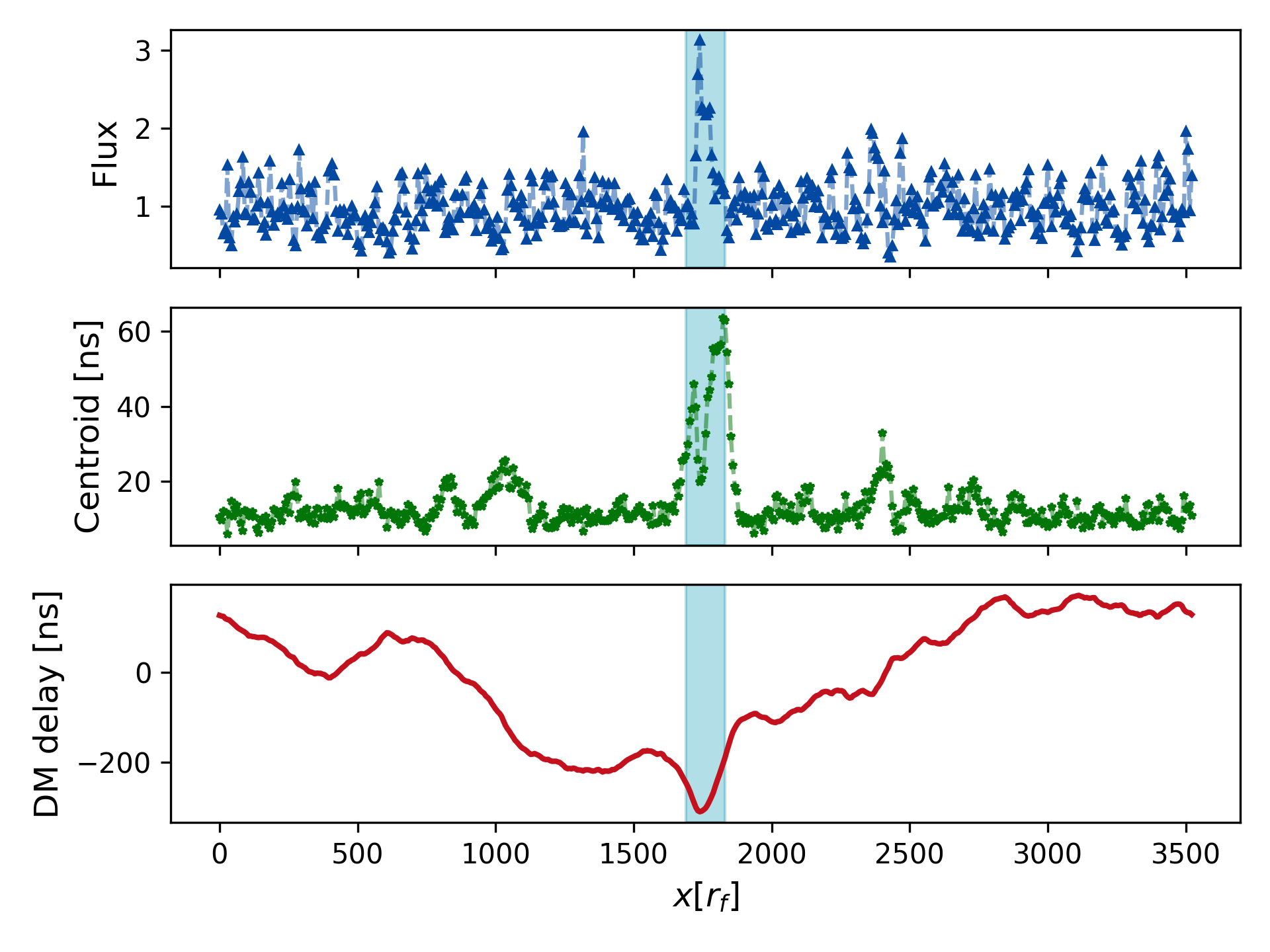} 
        \caption{}
        \label{subfig:iDM_centroid_flux_ESE}
    \end{subfigure} \\
    \caption{Simulations of isotropic scattering. The strength of scattering in these simulations is $ m_b^2 = 20$. \textbf{Panel a}: The bottom plot shows the DM variation from one instance of the isotropic phase screen. Equation \ref{eqn:DM_from_phases} was used to obtain DM values from a slice through the simulated phase screen at $L_{y}/2$ along the $x$ direction. The middle plot shows the measured centroid variation (scattering noise) and the top plot shows the associated flux density variation. \textbf{Panel b}: The main plot shows the average power spectrum of the DM variation and centroid variation signal obtained from ten instances of the phase screens. The best-fitting power law to the averaged power spectrum is given in the bottom left of the figure. The shaded region around each plot represents the scatter due to the stochastic nature of the phase variations. The dashed line in light brown shows a power law fit to DM variations with a spectral steepness of $\gamma=8/3$ as expected from a Kolmogorov medium. The inset shows the relationship between the centroid variation and the square of the gradient of the DM variation. \textbf{Panel C}: The plot shows the correlation between the centroid variations across the radio frequencies with respect to the center frequency of 1400 MHz. Different colors show the measurements obtained from ten instances of the phase screens \textbf{Panel D}: The plots show the DM, centroid and flux density variations when a Gaussian dip was added at the middle of the phase screen in order to simulate a region of under-density in the ISM. The shaded blue region represents one standard deviation of the Gaussian dip. }
    \label{fig:Simulations_results}
\end{figure*}

\begin{figure*}
    \centering
    \includegraphics[width=\linewidth]{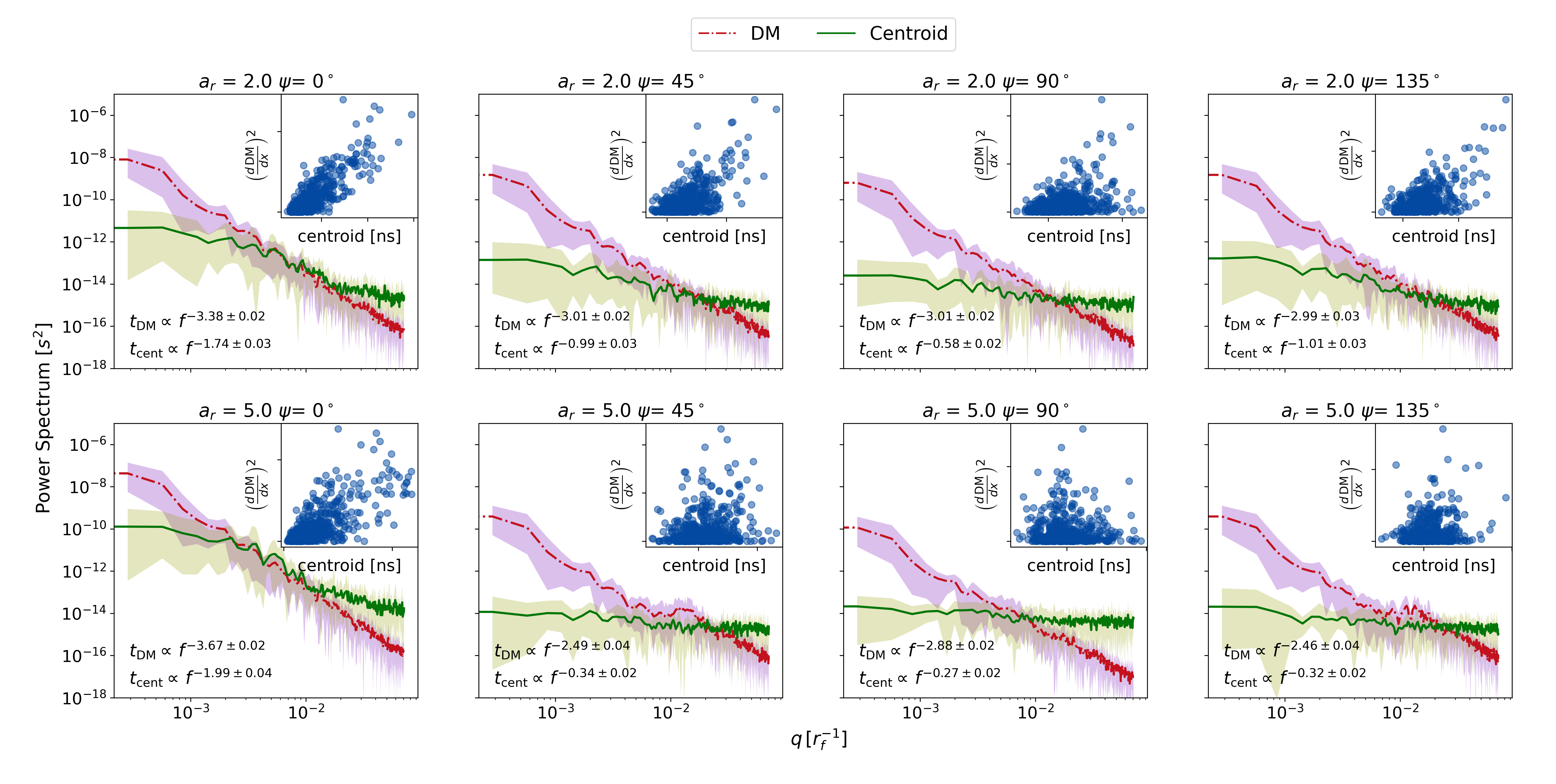}
    \caption{Averaged power spectra of chromatic processes for different levels and types of anisotropy. The best-fitting power law to the averaged power spectra in each case is given in the bottom left of each subplot. The shaded regions represent the scatter due to the stochastic nature of the phases. The strength of scattering is these simulations is $ m_b^2 = 20$. 
    The inner panel of each figure shows the scatter plot of the centroid variation with the square of the gradient of the DM variation.}
    \label{fig:PSD_correlations}
\end{figure*}

\begin{figure*}
    \centering
    \includegraphics[width=\linewidth]{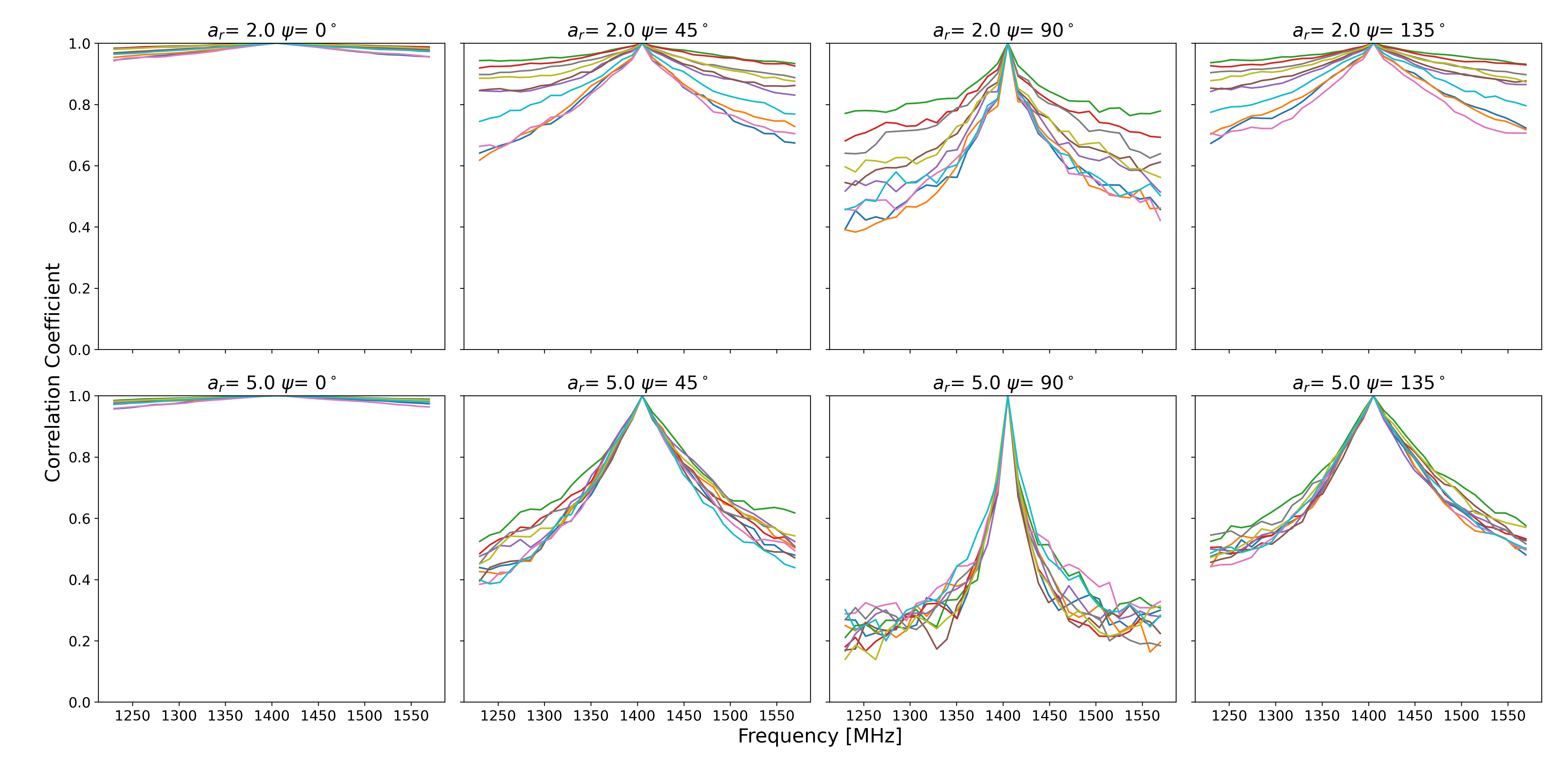}
    \caption{Correlation coefficient between the centroid variations measured at different frequency channels with respect to the center frequency ($1400$\,MHz). The strength of scattering in these simulations is $ m_b^2 = 20$. A significant decorrelation can be seen for orientation angles greater than $ \psi = 45^\circ$.}
    \label{fig:centroid_freq_corr_aniso}
\end{figure*}

\section{Results of simulations}
\label{sec:results_simulation}
We simulated $10$ realisations of phase screens for each of the scenarios considered. The top panel of Fig. \ref{fig:Dynamic spectrum} shows the full dynamic spectrum obtained for one instance of the isotropic phase screen. The bottom panel of Figure  \ref{fig:Dynamic spectrum} shows a portion of the screen that shows tilted structures in the dynamic spectrum, arising from the refractive phase gradients in the screen. The phase gradient displaces the apparent position of the source, which causes radio-frequency dependent displacement of the dynamic spectrum. 
We used the procedure described in section \ref{subsec:measurement_of_scattering_variations} to measure DM and centroid variations for each instance of phase screen. 
Fig. \ref{subfig:iDM_centroid_flux} shows the DM, centroid, and flux density time series for one instance in the case of isotropic scattering. The flux density variations and the centroid variations are both averaged over the entire simulation bandwidth. 
We computed the power spectrum from the time series corresponding to the DM and centroid variations for each instance of the phase screen. The power spectra from all instances were averaged together in order to reduce the fluctuation around the mean.
Fig. \ref{subfig:corr_chrom_sqdmdr_simulation} shows the averaged power spectrum of the DM and centroid variations. 
Further, we used the procedure outlined in Section \ref{subsec:Sub-banded scattering variations} to construct time series of centroid variations in $32$ frequency channels and these measurements were further used to obtain the correlations of centroid variations between different frequency channels. Fig. \ref{subfig:centroid_freq_corr_iso} shows the correlation coefficient with respect to the center frequency of $1400$\,MHz. Fig. \ref{subfig:iDM_centroid_flux_ESE} shows the DM, centroid and flux density time series obtained in the presence of a Gaussian converging lens superimposed on the underlying Kolmogorv turbulence. The blue shaded region shows the size of the lens. We followed a procedure similar to that applied in the case of isotropic scattering to analyse the simulations of anisotropic scattering as well. 
Fig. \ref{fig:PSD_correlations} shows the averaged power spectra, and Fig. \ref{fig:centroid_freq_corr_aniso} shows the centroid correlation coefficients from these anisotropic simulations. 
Here we summarise our findings:
\begin{enumerate}
    \item We identify a weak anticorrelation between the flux density variations and centroid variations from our simulations. We measured a Pearson correlation coefficient of $-0.27 \pm0.09$ for the flux density and centroid time series shown in Fig. \ref{subfig:iDM_centroid_flux}. A similar observation has also been reported in \citet{2010ApJ...717.1206C} for diffractive effects. In the theory of scattering, refractive effects from large scale phase variations have often been treated as perturbations to the diffractive scattering which arise from the small scale phase variations. For this purpose, \citet{1986ApJ...310..737C} and \citet{2017MNRAS.464.2075S} have defined refractive gain functions ($G$) in their analysis. In this framework, one may expect a positive correlation between flux density and arrival time variations, as both quantities are directly proportional to polynomials of $G$. Despite this, \citet{1995ASPC...72..357L} and \citet{1998A&A...334.1068L} have measured negative correlation from observations of PSR B1937$+$21. However, in a detailed analysis \citet{1998A&A...334.1068L} found that although a positive correlation between flux density and pulse arrival times is indeed expected, any left over DM delay in the arrival time measurements can make the overall measurement negative. As discussed earlier, a similar situation is present in our simulations. It is possible that a residual frequency-dependent DM delay is present in the centroid measurement, which might have resulted in its weak anticorrelation with the flux density variations.
   It is also possible that spatial averaging over a refractive scale is insufficient to remove the anti-correlation associated with diffractive effects.  We note that when we simulate large Gaussian lenses we observe positive correlation (see Figure \ref{subfig:iDM_centroid_flux_ESE}).  
    \item The power spectrum of the centroid variations is expected to be shallower than the power spectrum of the DM delay term. It can be noted from Fig. \ref{subfig:corr_chrom_sqdmdr_simulation} that the power spectrum at higher fluctuation frequencies is dominated by the centroid term rather than the DM term. It is also interesting to note that the strength and the spectral steepness of the centroid variation term depends on the orientation of the anisotropy with respect to the velocity direction of line of sight, as shown in Fig. \ref{fig:PSD_correlations}. The power spectrum flattens and approaches white noise when $\psi = 90^\circ$.
    \item In the case of isotropic scattering (Fig \ref{subfig:corr_chrom_sqdmdr_simulation}), high correlation of $0.72\pm0.05$ is found between the centroid variations and the square of the gradient of the DM variations from the direct line of sight. Theoretically, a linear relationship is expected as
    \begin{equation}
           \Delta t_{\rm ref} \propto \left ( \frac{d\,\rm DM}{dt} \right )^{2}.
           \label{eq:tref_propto_sqdmdr}
    \end{equation}
    A similar conclusion was reported earlier by \citet{1990ApJ...364..123F} for one dimensional screens and later by \citet{2010ApJ...717.1206C} and \citet{2017MNRAS.464.2075S} for two dimensional screens. These results can be explained on the basis of equation \ref{eq:t_ref}. The angle of refraction $\theta_{r}$ is proportional to the gradient of the phase variation in the IISM. As stated earlier, the DM at any epoch is proportional to the line of sight phase. The correlation is not expected to be 100\% as the perpendicular component (the component transverse to direction of motion) of the phase gradient is not sampled. Apart from that, the centroid variation measured in our work may contain contribution from the residual DM from the actual image center, which may reduce the correlation further.
    \item To account for scattering variations, PTA collaborations use the Gaussian process approach, which assumes the Gaussian nature of the arrival time variations. However, it can be seen from Fig. \ref{subfig:iDM_centroid_flux}, that the centroid variations show apparent non-Gaussianity with distributions skewed more towards higher delays. This can also be inferred from the relation \ref{eq:tref_propto_sqdmdr}, which suggests that the scattering delays would follow a $\chi^2$  distribution as they are proportional to the square of a Gaussian distributed DM delay\footnote{This assumes that the DM delays follow a Gaussian distribution.}. However, because of finite averaging over multiple PBFs, the distribution approaches the Gaussian limit according to the Central Limit Theorem. 
    \item As shown in Fig. \ref{fig:PSD_correlations}, a higher degree of correlation is observed between the centroid term and the square of the DM gradient term even for an anisotropic phase screen with orientation angle of $\psi = 0^\circ$ (anisotropy aligned in the direction of motion). However, the correlation gradually drops as the angle of orientation is moved away from the direction of velocity of line of sight. This suggests that measurements of scattering variations in PTA data sets can be used to assess the anisotropy of scattering.
    \item In the case of isotropic scattering, the centroid variations are highly correlated in frequency. However, it is interesting to note from Fig. \ref{fig:centroid_freq_corr_aniso} that for the anisotropic case the level of correlation between frequencies falls rapidly when the direction of anisotropy is moved away from direction of velocity of line of sight. The current models used in pulsar timing to describe the scattering noise assume that the noise is fully correlated across frequency. Our simulations show that this assumption fails in case of anisotropic scattering. This highlights the need to update the existing modelling strategy to incorporate the potential for spectral decorrelation in scattering noise. 
    \item We used the sub-banded measurements of the centroid variations to measure the chromaticity of the centroid variations. For this, a power law of the form $\tau(\nu) = A\nu^{-\alpha}$ was fitted to the mean of centroid variations in each channel, using the \textsc{curve\_fit} module of package \textsc{scipy}. 
    We obtained a value of $ \beta = 3.8\pm0.1$ which is smaller than the expected value of 4.4. 
    \item Large-scale regions of underdensities or overdensities in the IISM plasma can act as lenses for radio signals, causing a sudden increase or decrease in pulsar flux density. These events are known as extreme scattering events (ESEs) \citep{1987Natur.326..675F,2015ApJ...808..113C,2016Sci...351..354B,2018MNRAS.474.4637K}. 
    To simulate a region of plasma underdensity, we added a two-dimensional Gaussian phase structure in addition to the underlying Kolmogorov phase variations. Fig. \ref{subfig:iDM_centroid_flux_ESE} shows the results. The blue shaded region indicates the location of the simulated underdensity, which can also be identified as a decrease in the DM variation. The flux density at this location shows a peak because of the focusing effect, whereas the centroid variation shows two spikes coinciding with the rising and falling edges of the injected Gaussian. The centroid signal is in agreement with the equation \ref{eq:tref_propto_sqdmdr} showing that the relation can be a useful diagnostic for detecting large-scale structures in the IISM.
    
\end{enumerate} 

The results of our simulations provide predictions for the relationships between different observable signals originating from the scattering of radio waves in the IISM. These form the basis of our analysis of the MPTA data set, which we present in the next section. 


\section{Scattering noise in MPTA 4.5-year data}
\label{Sec:MPTA_Data}
The MPTA uses the MeerKAT radio telescope, which is situated in the Great Karoo region of South Africa. The telescope is a precursor to the upcoming SKA telescope and it currently operates with 64 dishes and with a low system temperature of $T_{\rm sys} \sim 18\,K$ \citep{2020PASA...37...28B}. Because of the large collecting area and low system temperature of the telescope, the MPTA can observe pulsars to higher precision with shorter integration times than the other Southern-hemisphere facilities. Capitalising on this, the MPTA observing strategy has been developed with the goal of maximising the array sensitivity to a GWB \cite[][]{2022PASA...39...27S}. The observations are made using the L-band receiving system, which has a central frequency $1283$ MHz and usable bandwidth of $\approx  780$\,MHz. The integration times for each pulsar have been set to achieve band-averaged uncertainty of $\rm 1\,\mu$s \citep{2022PASA...39...27S}. Due to the telescope's wide bandwidth and relatively high, fortnightly, observing cadence, the MPTA data set offers an excellent opportunity to study the long-term variations in the scattering properties of a large sample of millisecond pulsars spanning a range of lines of sight.

Recently, the MPTA collaboration produced a 4.5-year data set. The data set, comprising observations of 83 pulsars, has been used primarily for the search of a stochastic GWB \citep{2025MNRAS.536.1489M}. 
An essential part of a GWB search involves performing a single pulsar noise analysis (SPNA) to construct a noise model for each pulsar in the array.
This is done to identify various noise terms, apart from the GWB, that can be present in pulsar times of arrival (TOAs). A detailed noise analysis was performed to finalise the best suitable noise models for the MPTA 4.5-year data set. Here we give a summary of different stochastic and deterministic signals that were searched for in the noise analysis. The stochastic terms in PTA data sets are of two types: time-uncorrelated (white) noise and time-correlated (red) noise. MPTA uses parameters EFAC and EQUAD to model the white noise in the data by modifying the TOA uncertainties derived from the template matching techniques \cite[][]{J_H_Taylor_1993}. This modification is thought to be necessary because of systematic errors in pulse shapes due to sources such as radio frequency interference that are not captured by the template matching algorithms. The parameter ECORR was included to model the white noise correlated across radio frequencies at a single epoch. This could arise from processes such as pulse jitter \citep{2014MNRAS.443.1463S,2021MNRAS.502..407P}. The red noise in TOAs can arise due to various stochastic processes which are intrinsic and extrinsic to the pulsar. The red noise is further classified as achromatic and chromatic according to its radio-frequency dependence. 
The delays due to a red noise term can be expressed as Fourier series of the form 
\begin{equation}
    \Delta t_{\rm stoc} = a_{0} + \sum_{k=1}^{N}a_{k}\sin\left(\frac{2\pi}{T_{\rm obs}}kt\right) + \sum_{k=1}^{N}b_{k}\cos\left(\frac{2\pi}{T_{\rm obs}}kt\right),
    \label{eq:Fourier_series}
\end{equation}
where $T_{\rm obs}$ is the total observing span of the pulsar, $a_{k}$ and $b_{k}$ are the Fourier coefficients corresponding to a frequency component $k$, $N$ is the total number of Fourier coefficients. Although it is possible to fit for the Fourier coefficients, this is computationally expensive. Instead, it is usually assumed that the Fourier coefficients are Gaussian random variables whose second moments can be characterised by hyperparameters.
Under this assumption, it is possible to marginalise over the Fourier coefficients and only consider the hyperparameters. It is also common to assume that the Fourier coefficients follow a power-law such that the PSD is given by 
\begin{equation}
    p(f;A,\gamma,\beta) = \frac{A^2}{12\pi^2} 
    \left (\frac{f}{f_{c}} \right )^{-\gamma} 
    \left (\frac{\nu}{\nu_{c}} \right )^{-2\beta},
    \label{eq:red_noise_power_law}
\end{equation}
where amplitude $A$ and the spectral steepness $\gamma$ are the hyperparameters, $f$ is the frequency of the signal being modeled, $f_{c} = 1\rm yr^{-1}$ is the characteristic reference frequency, $\nu_{c}$ is a reference radio-frequency, chosen to be $\nu_{c} = 1400\,\rm MHz$. The parameter $\beta$ defines the chromaticity of the signal. 
For the MPTA noise analysis, four red noise terms were considered: achromatic red noise ($ \beta = 0$), DM variations due to IISM ($ \beta = 2$), stochastic component of the Solar wind ($ \beta = 2$, modelled according to \citet{2022ApJ...929...39H}), and scattering variations. The chromaticity of the scattering term depends upon the properties of the turbulence and can take different values along different lines of sight. Thus a scattering noise term with chromatic indices both fixed at $\beta=4$ and with $\beta$ as a free parameter was searched.

In addition to stochastic signals, MPTA noise analysis also included deterministic signals to account for discrete structures in TOAs. For example, the IISM can contain discrete structures produced by a turbulent process. Propagation of radio waves through these structures could manifest in arrival time variations that are not well modeled by the power-law processes described in eq. \ref{eq:red_noise_power_law}.
An alternate approach is to model the variations that has a Gaussian functional form as described in \citet{2023ApJ...951L...7R}: 
\begin{equation}
    \Delta t_{\rm Gauss} = A_g \rm exp \left(
    \frac{(t-t_{g,0})^2}{2\sigma_{g}^2} 
    \right) \times
    \left( \frac{\nu}{\nu_{c}}\right)^{-\beta_{g}},
    \label{eq:Gaussian_bump}
\end{equation}
where $A_{g}$ is the peak amplitude  of the Gaussian perturbation, $t_{g,0}$ is the epoch associated with the centre of the Gaussian, $ \sigma_{g}$ is the width and $\beta_{g}$ is the chromaticity. Apart from that, the motion of Earth around the Sun can give rise to deterministic annual chromatic variations \citep{2013MNRAS.429.2161K}. For example, annual DM variations would be expected if the pulsar Earth line of sight revisits the same path through the ISM. Following from 
\citet{2021MNRAS.502..478G}, these perturbations are modeled as 
\begin{equation}
    \Delta t_{\rm annual} = A_{s} \sin(2\pi t \times f_{\rm yr} + \phi ) \times 
    \left( \frac{\nu}{\nu_{c}}\right)^{-\beta_{s}},
    \label{eq:Chrom_annual}
\end{equation}
where $A_{s}$ is the amplitude of the sinusoid, $\phi$ is the phase of the signal, $\beta_{g}$ is the chromaticity and $f_{\rm yr}$ is the frequency of a year. The chromaticity of both these signals have also been treated as a free parameter during the MPTA 4.5-year pulsar noise analysis. 

Lastly, the presence of solar wind causes extra delays in the pulse arrival times when the Earth-pulsar line of sight is in the proximity of the Sun, To model these, MPTA 4.5-year noise analysis used a spherically symmetric model for the electron density around the sun parameterised by $n_\oplus$, which describes the solar wind electron density at 1 AU \citep{2006MNRAS.369..655H}.

Table \ref{tab:MPTA_noise_params} lists the types of signals along with their corresponding parameters which were searched in the SPNA of the pulsars. A noise model for each pulsar was constructed based on the Bayesian evidence comparison for the inclusion of each noise term. A detailed discussion on the methods and the resulting noise models is given in \citet{2025MNRAS.536.1467M}. Here we summarise the results of the noise analysis with a focus on the chromatic signals pertaining to IISM. Out of 83 MSPs, 49 pulsars had detection of DM noise, and 23 pulsars had detection of scattering noise. The chromatic index of the scattering noise was in the range of $2.5 \le \beta \le 8.4$, with 10 pulsars favouring a fixed chromatic index of $\beta = 4$. Out of these 23 pulsars, only two had a $\beta<4$. One of them was PSR~J1825$-$0319, which had a chromatic index of $\beta = 2.5^{+0.37}_{-0.18}$, close to but inconsistent with DM noise. The second pulsar, J1652$-$4838 had a chromatic index of $\beta=2.98^{+0.57}_{-0.17}$, more consistent with scattering indices observed for pulsars \citep{2004ApJ...605..759B}. 
Gaussian chromatic events were detected in 15 pulsars with measured chromatic indices in the range $0.52\le \beta_{g} \le 7.4$. Similarly, 8 pulsars had detection of chromatic annual variations whose chromatic indices were in the range $1.5\le \beta_{s}\le 5.04$. Table 2 of \citep{2025MNRAS.536.1467M} lists the maximum a-posteriori (MAP) values along with the confidence intervals on each parameter. It can be noted that for most of the pulsars the properties of the deterministic chromatic processes were not well constrained.

\begin{table}
    \centering
    \begin{tabular}{m{0.25\linewidth} m{0.25\linewidth} m{0.35\linewidth}}
        \toprule
        Type                  & \multicolumn{2}{l}{Parameters} \\
        \midrule
        White Noise           &                   &  $\rm  EFAC$, $\rm EQUAD$, $\rm  ECORR$ \\
        \midrule
        \multirow{2}{4em}{Red noise (achromatic)}& Spin noise       & $\rm log_{10}A_{ red}$, $\gamma_{\rm red}$ \\
        \cmidrule{2-3}
                              & GWB              & $\rm log_{10}A_{\rm gw}$, $\gamma_{gw}$ \\
        \midrule
        \multirow{3}{4em}{Red noise (chromatic)} & DM noise         &  $\rm log_{10}A_{ DM}$, $ \gamma_{\rm DM}$\\
        \cmidrule{2-3}
                              & Chromatic noise & $ \rm log_{10}A_{ chrom}$, $ \gamma_{\rm chrom}$, $ \beta$ \\
        \cmidrule{2-3}
                              & Stochastic solar wind & $\rm log_{10}A_{ sw}$, $\gamma_{\rm sw}$ \\
        \midrule
        \multirow{2}{4em}{Other chromatic signals}     & Annual chromatic variations    & $ \rm log_{10}A_{s}$, $\beta_{s}$, $\phi$\\
        \cmidrule{2-3}
                              & Chromatic Gaussian Events       & $ \rm log_{10}A_{g}$, $ \beta_{g}$, $ t_{\rm g,0}$, $ \sigma_{g}$, $\rm Sign[+/-]$\\ 
        \cmidrule{2-3}
                              & Deterministic solar wind & $ n_{\oplus} (\rm cm^{-3})$ \\
        \bottomrule
    \end{tabular}
    \caption{MPTA noise models and parameters.}
    \label{tab:MPTA_noise_params}
\end{table}


\subsection{Methodology}
To assess the effectiveness of the current IISM noise modelling techniques, we analysed the time-domain realisations of various noise processes for each pulsar. The flux density variations of each pulsar were compared with these time series of noise processes. For deterministic signals, we used the MAP values of the parameters to construct time series using equations \ref{eq:Gaussian_bump} and \ref{eq:Chrom_annual}. For stochastic signals, we used the MAP values of the hyperparameters to solve for the Fourier coefficients in equation \ref{eq:Fourier_series} and thereby obtained a maximum-likelihood realisation of DM, scattering noise and stochastic solar wind processes. The stochastic solar wind noise realisations were constructed using the model presented in \citet{2022ApJ...929...39H}, which models deviations in TOAs arising from stochastic perturbations on the mean solar wind electron density. The pulsar timing software package \textsc{pint} \citep{2021ApJ...911...45L} was used to perform the fit. We created the time realisations of all signals at a reference frequency of $\nu_{c} = $1400\,MHz, by scaling the multi-frequency noise realisations using the corresponding values of the chromaticity parameter $\beta$. The stochastic DM noise variations and the scattering noise variations obtained in this way are referred to as `DMGP' and `CHROM' in this work. 

The deterministic signals were further scrutinised before being included in the following analysis. We performed a visual inspection of the constructed time series of the deterministic signals. It showed that in some cases the deterministic signals were insignificant and inconsistent. For instance, out of 15 pulsars, 8 had an unusually long event duration of much more than one year for the chromatic Gaussian event. The flux and DM-epoch time series of these pulsars did not show any noticeable structure consistent with the Gaussian event. The reason for this unusual behaviour is not understood and requires further study. In other cases, the chromaticity of the signal was not well constrained and their amplitude were negligible relative to the white noise in the data. For this reason, the contribution of the deterministic signals was neglected in many pulsars except for the following six pulsars. The Gaussian events in PSR J1421$-$4409 and PSR J1918$-$0642 and the chromatic annual variations in PSR J1045$-$4509 and PSR J1804-2858 were included in our analysis. The chromaticity of these signals was found to be more consistent with the scattering noise, and hence they were added to the CHROM time series of these pulsars. The Gaussian event in PSR J1737$-$0811 and the chromatic annual variation in PSR J1643$-$1224 were also included in this analysis, but since the chromaticity of these signals was more consistent with that of DM noise, they were treated as DM variations and combined with the DMGP time series of these pulsars. Although additional dispersion due to stochastic and deterministic solar wind signals are both chromatic in nature, here we focus on signals that arise in the IISM.  


Another method to obtain the DM time delay variation is to simply fit for a $\nu^{-2}$ term to the residuals of each epoch separately. A recent study by \citet{2024A&A...692A.170I} has shown that DM measurements using this approach are less impacted by achromatic red noise in timing residuals. However, this method ignores the fact that the arrival times may contain additional scattering delays which scale differently than $\nu^{-2}$.
It is also sub-optimal and results in larger DM uncertainties \cite[][]{2013MNRAS.429.2161K}. 
Although the obtained DM time series may be less precise, it can still be useful to corroborate the DMGP time series obtained via the Gaussian processes approach. Thus, we used the pulsar timing software \textsc{tempo2} \citep{2006MNRAS.369..655H} to fit for DM using TOAs from each epoch independently. For this fit, all other timing model parameters were held fixed and the post-fit DM values were used to calculate the delays at the reference frequency of $\rm 1400\,MHz$. This time series is referred to as `DM-epoch'.  

The flux density variations of the pulsars contain information about the characteristics of propagation in the IISM. For our analysis, we used the flux densities measured from the custom processing pipeline, \textsc{meerpipe}. The pipeline performs the flux calibration and data reduction of all MPTA observations. 
We then obtained a flux measurement averaged over the entire observing bandwidth of $\rm \approx 780 \,MHz$. The time series was then compared to that of other noise signals. 

\subsection{Results}
\label{subsec:Results}
\begin{figure*}
    \centering
    \includegraphics[width=\linewidth]{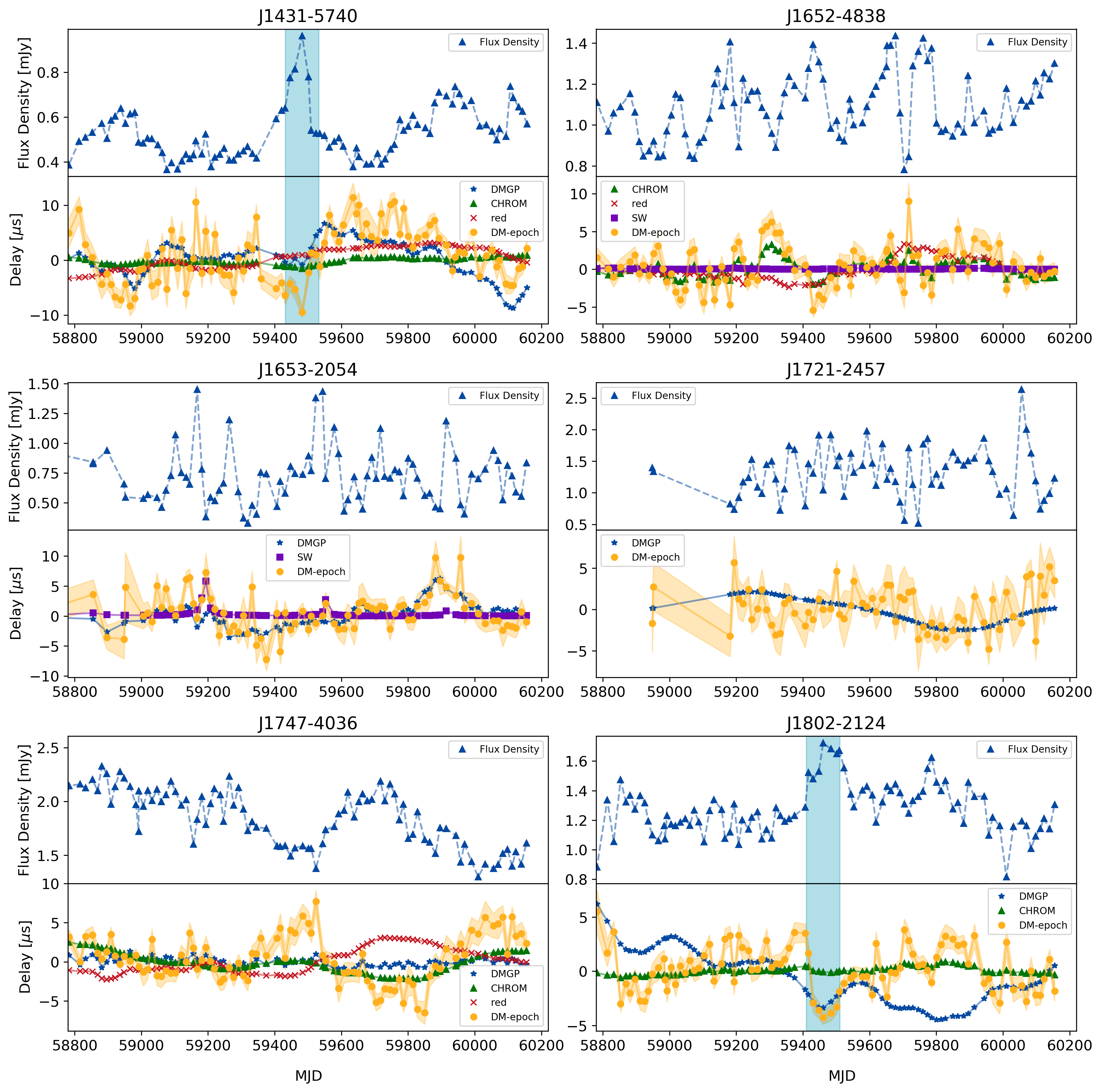}
    \caption{Time series of noise processes and flux densities for a sample of MPTA pulsars. The top panels of all plots show the band-averaged  flux density variations. The corresponding time series of different noise processes are shown in the bottom panels. These are obtained from the noise analysis of MPTA data set and are referenced to $\rm 1400\,MHz$. The yellow shaded region around the DM-epoch time series represents the error region based on the DM uncertainties reported by \textsc{tempo2}. The blue shaded region in the plots for PSR J1431$-$5740 and PSR J1802$-$2124 represents the underdensity in the IISM.}
    \label{fig:Noise_signals}
\end{figure*}

The statistical analysis presented in Fig. 7 of \citet{2025MNRAS.536.1467M} shows that the amplitude of the chromatic noise signal and the DM noise signal are positively correlated with the DM of a pulsar. This supports the IISM origin of the chromatic noise signals, as it is well known that the radio waves from higher DM pulsars are more strongly scattered because of propagation through a greater number of screens \citep{1991Natur.354..121C}. Therefore, these pulsars are expected to show a stronger chromatic noise. The time series analysis presented in this work provides additional insights into the nature of these signals. We analysed the time series of all MPTA pulsars.
In Fig. \ref{fig:Noise_signals}, we highlight six pulsars that illustrate the key findings of our analysis. The upper panels in these plots show the flux density variations and the lower panels show time series of noise signals. 
The figure also shows time series of signals like solar wind and achromatic red noise in the pulsar residuals, when those signals were found to be present.  

\begin{figure}
    \centering
    \includegraphics[width=1\linewidth]{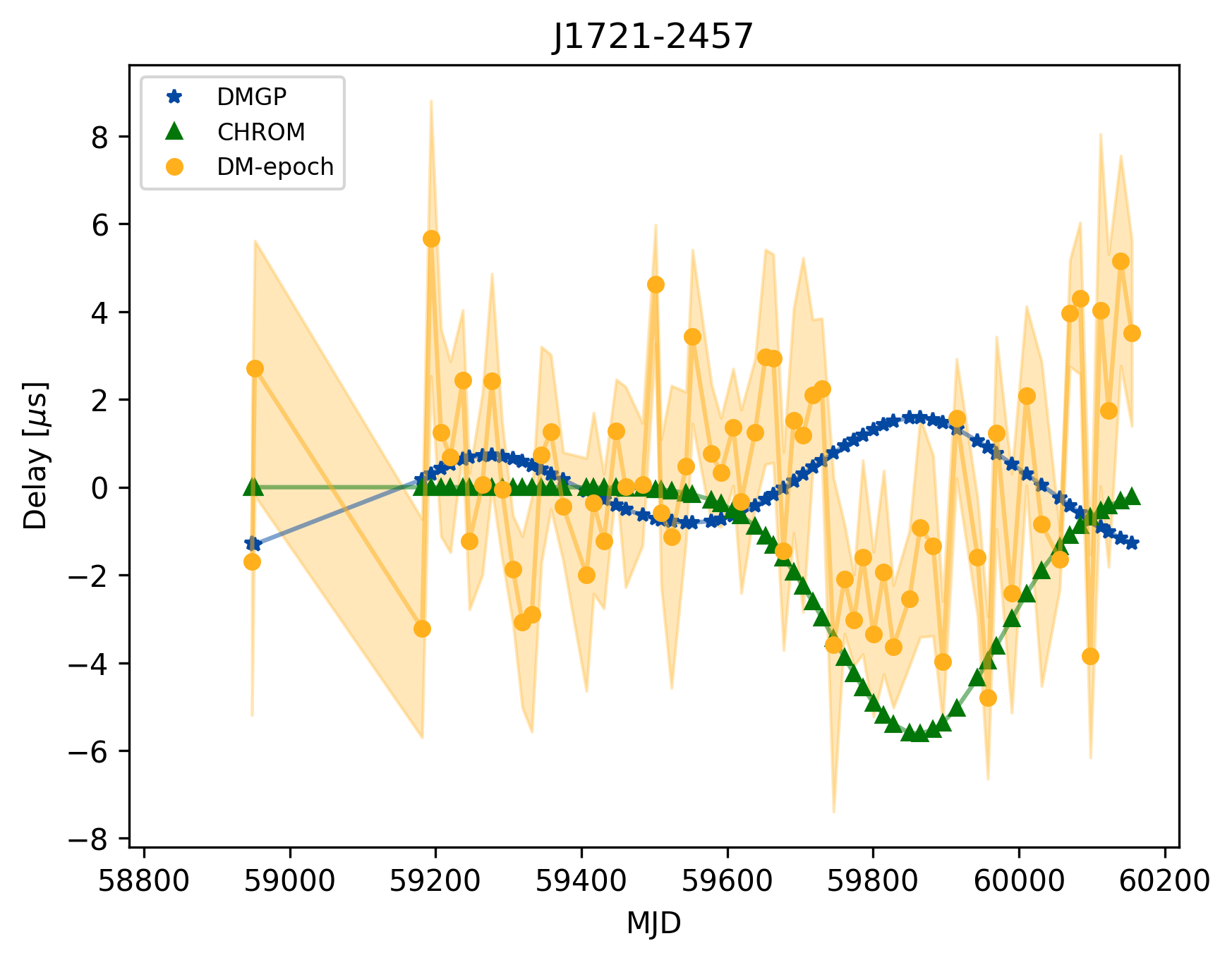}
    \caption{Noise realisations for PSR~J1721-2547 based on a preliminary noise models. Here the CHROM time series represents a chromatic Gaussian event with chromaticity of $\beta_{g} = 2.09$ and DMGP represents the DM variations. A strong anticorrelation can be noticed between the DMGP and CHROM at epoch 59850.}
    \label{fig:J1721-2457_17April_noise_realisations}
\end{figure}

\begin{figure*}
    \centering
    \includegraphics[width=\linewidth]{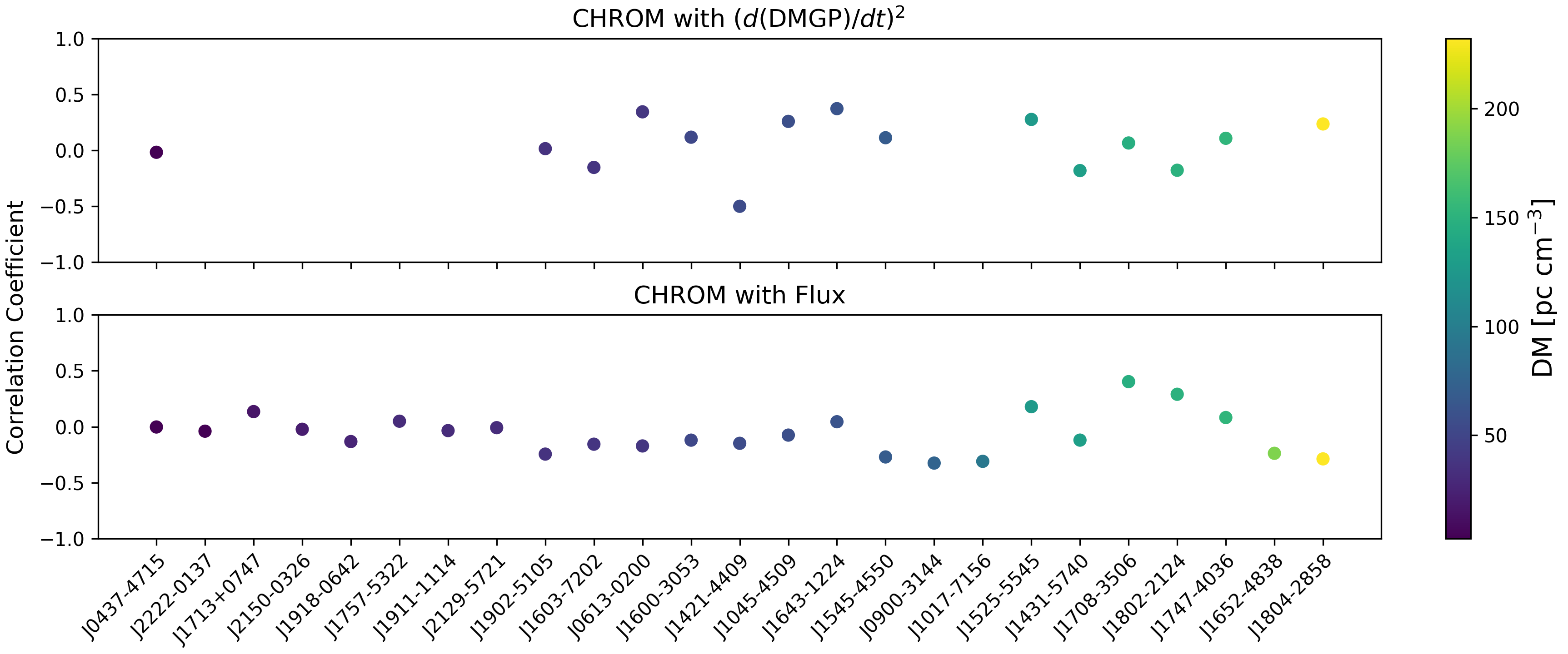}
    
    \caption{Correlations between noise processes and flux densities for the MPTA pulsars. The upper panel shows the measured correlation between CHROM and the square of the DMGP derivative and the lower panel shows the measured correlation between CHROM and flux density variations from the real MPTA data. The color of the circles represents DM of the pulsar.}
    \label{fig:MPTA_chrom_correlations}
\end{figure*}

\begin{figure}
    \centering
    \includegraphics[width=1\linewidth]{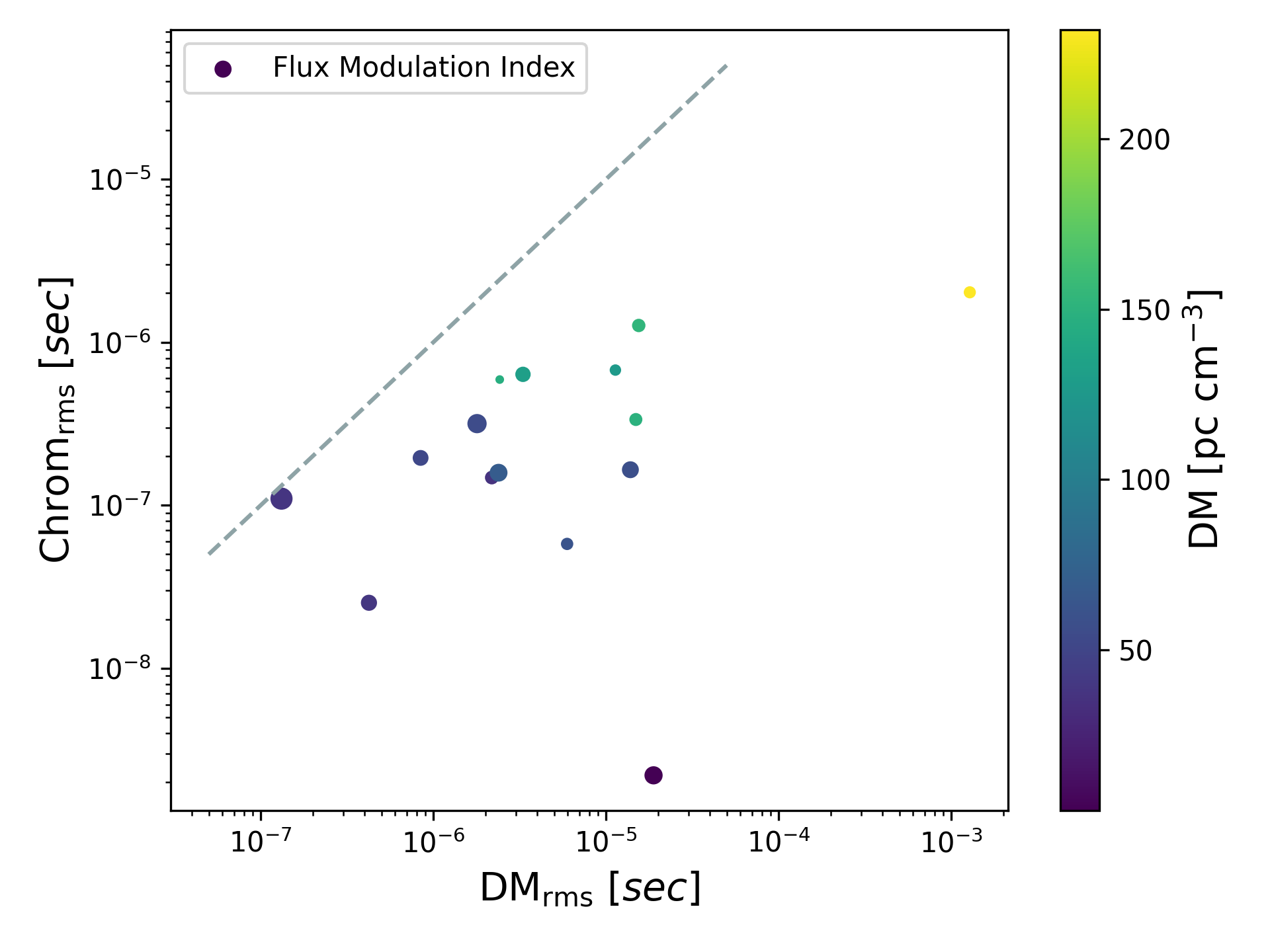} 
    \caption{ Strength of scattering and DM variations. Here the DM variations include the contributions from the post fit linear ($\dot{\rm DM}$) and quadratic ($\ddot{\rm DM}$) terms along with the DMGP term from noise analysis. The sizes of the circles represents the flux density modulation index and the colors of the circles represent the DM of the pulsar. The dashed line represents equal strength of scattering and DM variations.
    }
    \label{fig:Chrom_DM_rms}
\end{figure}

From this investigation, we note the following:
\begin{enumerate}


    \item For the majority of MPTA pulsars the DM-epoch time series correlates well with the DMGP time series. For instance, Fig. \ref{fig:Noise_signals} shows the time series for PSR J1653$-$2054. In this case $\approx 60\%$ of the DMGP measurements were found to lie within the one sigma error region of the DM-epoch time series.

    \item However, in some cases the DMGP time series shows deviation from the DM-epoch time series. The SPNA of such pulsars also favour addition of chromatic noise terms to the noise model. This indicates that the measurement of DM-epoch time series could be biased and that the residuals may contain an additional frequency-dependent term, potentially due to scattering. However, certain features in the time series of these processes suggest that sometimes the scattering noise models influence the fitting of other frequency-dependent signals like DMGP. 
    For instance, in case of PSR J1802$-$2124, a clear anticorrelation between CHROM and DMGP is seen from MJD 59600 to 60000.  
    To further illustrate the anticorrelations between noise signals, in Fig. \ref{fig:J1721-2457_17April_noise_realisations} we show the noise realisations of PSR J1721-2457 using a noise model obtained at a preliminary stage of MPTA noise analysis. For this pulsar, a loud Gaussian deterministic event with a chromaticity of $\beta_{g} = 2.09$ was thought to be present along with DM noise. It can be clearly seen that the Gaussian event appearing at MJD 59850 is anticorrelated with a crest in the DMGP time series. However, the DM-epoch time series does not show any noticeable structure. The noise realisations from the finalised noise model for this pulsar are shown in Fig. \ref{fig:Noise_signals}. In this case, a single DMGP time series was found sufficient to account for the DM variations in the data. This shows that the apparent anticorrelations seen between DMGP and CHROM may also arise as a result of the artefact of the fitting process rather than from an astrophysical origin. 
    

    \item  A comparison of chromatic processes and flux densities indicates a potential presence of scattering events along the lines of sight of pulsars PSR J1431$-$5740 and PSR J1802$-$2124. The regions shaded in blue in Fig. \ref{fig:Noise_signals} show the epochs at which rise in flux density and a dip in the DM-epoch and DMGP was seen. It is possible that CHROM, because of its covariance with DMGP, is absorbing some of the DMGP signal at these epochs. However the co-occurrence of enhanced flux density and dips in both DMGP and DM-epoch time series can be an indication of lensing effect in the IISM. One possible explanation could be the presence of plasma underdensity along these sight lines, which acts like a lens and causes refractive focusing of the flux. Such events can also cause significant changes in the measured scintillation bandwidth $\Delta \nu_{d}$ at those epochs \citep{2015ApJ...808..113C,2018MNRAS.474.4637K}. This is because the scatter broadening and scintillation are the manifestations of the same underlying process which results in relation $\Delta \nu_{d} \propto 1/\tau_{s}$. However, because of the shorter time span and a coarser frequency resolution of the MPTA observations (856 kHz), we were unable to measure the scintillation bandwidths for these pulsars.


    \item For PSR J1652$-$4838, a single stochastic scattering noise process was sufficient to adequately account for frequency dependent noise in the residuals. The chromaticity of this term was found to be $\beta = 2.98^{+0.57}_{-0.17}$ \citep{2025MNRAS.536.1467M}. For this pulsar, instead of DM variations, a stochastic solar wind was detected whose time series is shown in Fig. \ref{fig:Noise_signals}. The observed dominance of scattering delay terms is in contradiction with the conclusions of previous studies and the results of our simulations. 


    \item The results of our simulations suggest that the square of the gradient of the DM variations can have strong correlations with the scattering noise signals. Motivated by this, we have measured the Pearson correlation coefficient between the squared gradient of the DMGP time series and CHROM time series. The results are displayed in the upper panel of Fig. \ref{fig:MPTA_chrom_correlations}. 
    The data neither show a consistent trend among all the pulsars nor any significant correlation measurement for an individual pulsar. Similarly the bottom panel of Fig. \ref{fig:MPTA_chrom_correlations} shows the measured correlations between CHROM time series and the flux density variations. We do not find anticorrelations in the data as predicted by the simulations.

    \item Fig. \ref{fig:Chrom_DM_rms} shows a scatter plot of the relative strengths of the scattering variations and DM variations. Here the scattering variations can be adequately represented by the CHROM time series.  For the DM variations we have added the contributions of the linear ($\dot{\rm DM}$) and quadratic ($\ddot{\rm DM}$) terms of the Taylor series representation to the stochastic DMGP term. We quantify the signal strength by its root mean square (RMS) value. This is done to minimise the uncertainty arising from the high covariance between the parameters $A_{\rm chrom}$, $\gamma_{\rm chrom}$ and the chromaticity $\beta$. The circles in the scatter plot are color coded according to the DM of the pulsar. The sizes of the circles represents the flux density modulation index of a pulsar, which is the ratio of mean of the flux density variation to its standard deviation. As shown in the plot, the pulsars with higher DM exhibit a lower flux density modulation index as they experience higher scattering. The dashed line demarcates the regions of DM noise and scattering noise dominance. It can be seen that for all pulsars the scattering variations are approximately an order of magnitude weaker than the DM variations. 
\end{enumerate}

\section{Understanding the unusual chromaticity of chromatic noise}
\label{Sec:Chromaticity}

Theoretical investigations and simulations have predicted chromaticity for the scattering noise as high as  $\beta \approx 6.4$ \citep{2017MNRAS.464.2075S}, but not higher. 
However the measurements of $\beta$ from the multi radio frequency observations of pulsars can be complicated by factors like inherent radio frequency evolution of the pulse profile or inaccuracies in the assumed PBF for a line of sight. For example, a recent study by \citet{2025ApJ...986..191G} demonstrates that a mismatch between the true and assumed PBFs or pulse widths of a pulsar can significantly affect the $\beta$ measurements, with $\beta > 6.4$ in some cases. Additionally, pulsars can exhibit significant frequency evolution in their pulse profiles in the radio band, which, if uncorrected, can induce systematic errors in the estimation of radio frequency dependent effects in timing \citep{2014ApJ...790...93P,2015ApJ...813...65N}. For this reason, frequency dependent templates (two-dimensional templates) of pulsars can be used to measure the arrival times of pulses \citep{2019ApJ...871...34P}. The MPTA 4.5-year data set was created using two-dimensional templates in 32 frequency channels at L-band. However, noise analysis of this data set has found chromaticity of $\beta  = 7.95^{+1.41}_{-0.67}$ for PSR J0437$-$4715 and $\beta  = 8.83^{+1.96}_{-1.15}$ for PSR J1643$-$1224. The reason for these unexpected values is not clear and warrants further investigation. One possible explanation of this could be the presence of unmodelled frequency evolution in the two-dimensional templates. In this part of the work, we show another possible origin of high chromaticity which is independent of the complexities of scattering studies from pulsar observations, but arising from the interplay between noise models when modelling timing delays originating in the IISM. 
We use our multi-frequency IISM simulations to demonstrate the effect. Our method of measuring centroid variations of sub-banded PBFs from the simulations allows us to construct DM and scattering time series at different frequency channels. It also allows us to measure the expected chromaticity of scattering variations. We then check for the correct recovery of the chromaticity when the contemporary noise modelling techniques are used. 

\begin{figure*}
    \begin{subfigure}[t]{.49\linewidth}
        \centering\includegraphics[width=1\linewidth]{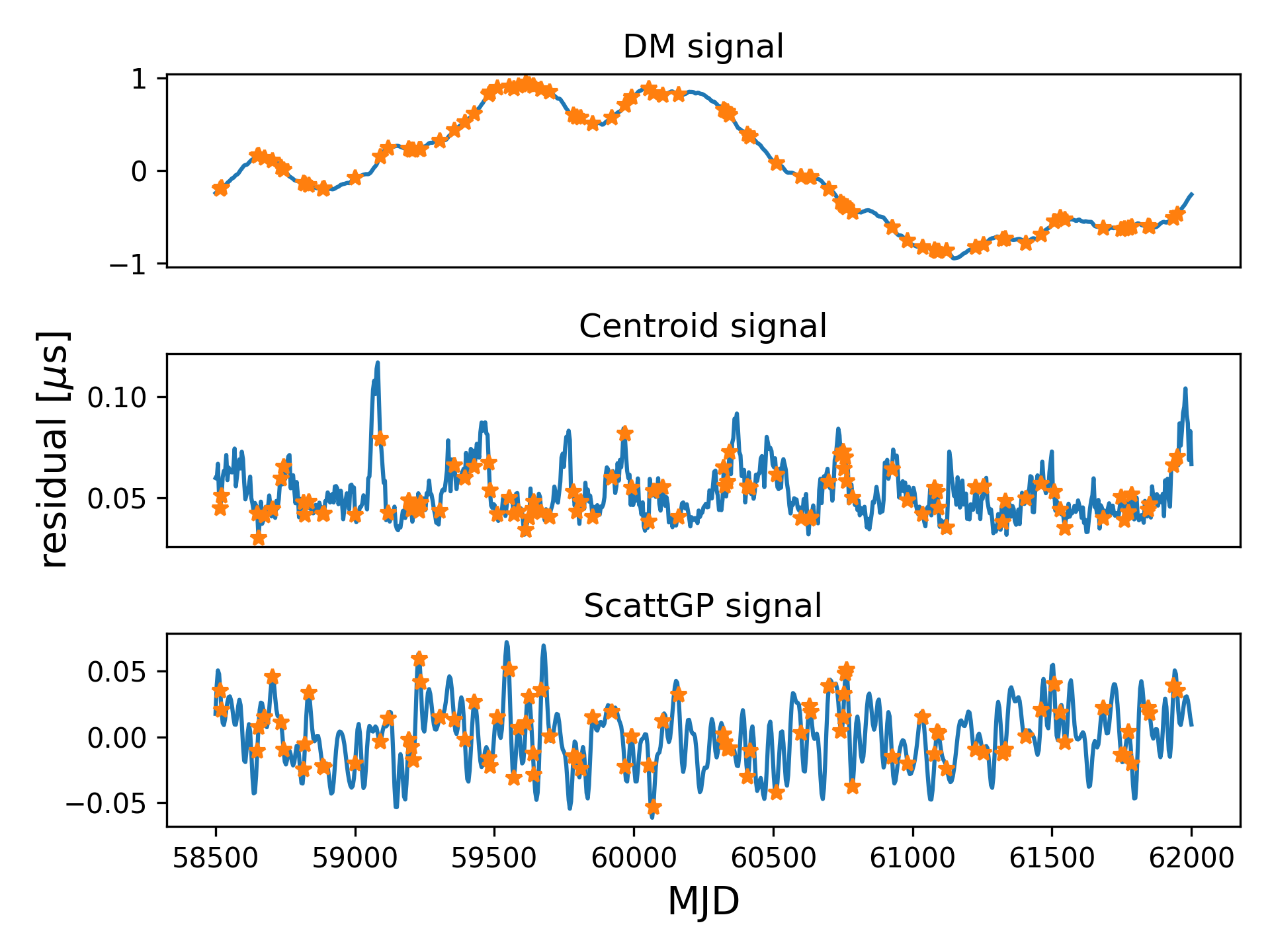} 
        \caption{}
        \label{subfig:TOAs_mb280_injection}
    \end{subfigure}
    \begin{subfigure}[t]{.49\linewidth}
        \centering\includegraphics[width=1\linewidth]{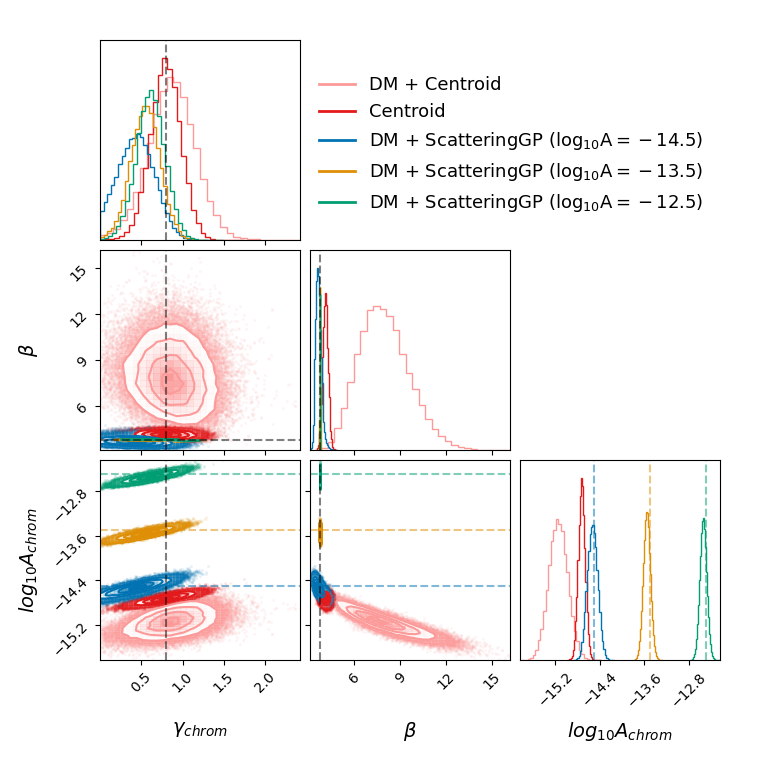}
        \caption{
        }
        \label{subfig:corner_dmchrom_compare}
    \end{subfigure}\\
    \caption{\textbf{Panel a}: Simulated DM and centroid variations.  The simulations were underertaken with the \textsc{scintools} code and the variations are referenced  to a centre frequency of $1400\,\rm MHz$. The bottom plot shows a signal simulated using the Gaussian process model. The orange stars show the epochs at which the signals were sampled.  \textbf{Panel b}: Recovered values of the chromatic noise parameters. The injected value of chromaticity of the centroid variations was $\beta = 3.8$, based on our multi frequency IISM simulations. The dotted lines show the injection values of the ScatteringGP signal.
    }
    \label{fig:Unusual_Chromaticity}
\end{figure*}

For this purpose we ran a separate isotropic simulation with a higher scattering strength of $ m_b^{2} = 80$ for a length spanning $\sim 500 \,s_{\rm ref}$. This simulation was centered at $1400$\,MHz with a bandwidth of $350$\,MHz. We followed the procedure mentioned in section \ref{subsec:Sub-banded scattering variations} to generate the sub-banded PBFs at $32$ frequency channels. After averaging for $\sim 0.5\,s_{\rm ref}$, the PBFs were used to measure the DM and centroid variations from this simulation. The top panel and the middle panel of Fig. \ref{subfig:TOAs_mb280_injection} show the simulated DM and centroid variations at 1400 MHz.  Then using the recent ephemeris of the PSR J1643$-$1224, we simulated ideal TOAs for a duration of 4.5 years starting from MJD 58500. The ideal TOAs were generated in $32$ frequency channels covering a bandwidth of $350$\,MHz centered at $1400$\,MHz to align with the setup of $ m_b^{2} = 80$ simulation described in this section. As J1643$-$1224 is a moderately high DM pulsar, its scattering strength can be much higher than in our simulation, however our simulation could still demonstrate a possible origin of the high chromaticity. To imitate the irregular sampling of TOAs in real data, we randomly sampled the DM and centroid signals at 90 epochs (shown as stars in Fig. \ref{subfig:TOAs_mb280_injection}), such that the average cadence of observation was $\sim 14$ days. The sampled values were used for injections into the simulated residuals as needed at their corresponding epochs. The precision of the TOAs was set to 5\,ns so that the scattering noise is sufficiently larger than the white noise level and hence should be detectable.

Using this method, we first created two simulated pulsar timing data sets, one containing only the DM signal and the other containing only the centroid signal. This was done to test the recovery of both signals independently from each other. Then, for the third data set, both DM and centroid signals were injected into the ideal TOAs. This allowed us to test the impact on the recovery of both signals when present together in residuals. As noted in section \ref{sec:results_simulation}, the centroid signal can be non-Gaussian in nature. To study the effect of non-Gaussianity on parameter estimation, we also conducted additional tests in which the centroid signal was replaced with another Gaussian distributed simulated signal, similar in characteristics to the centroid signal. This signal is called `ScatteringGP' and it was generated based of the current Gaussian process model of scattering noise. We used the values of $\gamma$ and $\beta$ from table \ref{tab:unusual_chromaticity}, obtained from the `centroid only' search to simulate this new signal. In addition, we created three instances of ScatteringGP but with increasing amplitudes of $\rm log_{10}A = -14.5,-13.5,-12.5$. The bottom panel of Fig. \ref{subfig:TOAs_mb280_injection} shows an example with the lowest amplitude $\rm log_{10}A = -14.5$, which is comparable to the centroid signal. The ScatteringGP signal was sampled at the same 90 epochs and injected into the ideal TOAs along with the DM signal.

Standard PTA noise analysis was conducted on these data sets using the software \textsc{enterprise} \citep{2020zndo...4059815E}. The DM variations were searched using a Gaussian red noise process with a fixed chromatic index of $\beta = 2$ where as the centroid variations were searched using an another Gaussian process (CHROMGP) with a variable chromaticity parameter. The white noise in the data was modelled using the parameter EFAC. Table \ref{tab:unusual_chromaticity} lists the values of parameters recovered in all the tests and Fig. \ref{subfig:corner_dmchrom_compare} shows the posteriors of the CHROMGP. 
In the absence of DM signal in the residuals, the Gaussian process approach with a power-law spectral model is able to recover the essential features of the centroid signal. For example, the recovered chromaticity was found to be $\approx 4$, which matches closely with the expected value of $3.8$. But in the presence of DM signal, the chromaticity of the centroid signal peaks at $\beta = 7.8$.  We note that there is significant covariance between $\beta$  and amplitude of the chromatic process. It is also important to note that the EFAC parameter was measured in excess by $\approx 25\%$ when the centroid signal was used in these tests. This excess EFAC could have arisen due to non-Gaussianity of the centroid signal, or it could be a result of the finite scintle effect which gives rise to additional white noise in the data. 
In contrast, from the analysis of ScatteringGP signal, we were able to correctly recover all the noise parameters in all cases. Additionally, the posterior distributions for the chromatic process parameters are also more tightly constrained than the centroid case. This implies that, in an effort to adapt to the non-Gaussian centroid signal, the Gaussian process model can make compromises and produce unintended results for the scattering measurements. However, when scattering noise is injected following a Gaussian process (ScatteringGP), the modelling works as expected.

These results indicate that the measurement of the steep chromatic index for the scattering noise term can indeed be caused by the IISM due to misspecification of scattering noise. If the noise is not a Gaussian process, steeper chromatic indices can be induced because of complex interplay between the DM and chromatic noise models. PTA collaborations may need to adopt a better modelling strategy to solve this problem.

 \begin{table*}
    \begin{center}
        \begin{tabular}{lcccccc}
            \toprule
            Signal injection & EFAC & $ \gamma_{\rm chrom}$ & $ \rm log_{10}A_{ chrom}$ & $ \beta$ & $ \gamma_{\rm DM}$ & $ \rm log_{10}A_{ DM}$ \\ 
            \midrule
            DM & $0.99^{+0.01}_{-0.01}$ & -- & -- & -- & $3.16^{+0.16}_{-0.16}$ & $-13.70^{+0.04}_{-0.04}$ \\ \\
            Centroid & $1.25^{+0.02}_{-0.02}$ & $0.79^{+0.18}_{-0.18}$ & $-14.73^{+0.06}_{-0.06}$ & $4.13^{+0.17}_{-0.17}$ & -- & -- \\ \\
            DM + Centroid & $1.21^{+0.02}_{-0.02}$ & $0.87^{+0.27}_{-0.28}$ & $-15.15^{+0.17}_{-0.17}$ & $7.83^{+1.76}_{-1.50}$ & $3.01^{+0.16}_{-0.15}$ & $-13.69^{+0.04}_{-0.04}$ \\ \\
            \midrule
            DM + ScatteringGP ($\rm log_{10}A = -14.5$) & $0.98^{+0.01}_{-0.01}$ & $0.44^{+0.22}_{-0.23}$ & $-14.53^{+0.09}_{-0.09}$ & $3.66^{+0.20}_{-0.16}$ & $3.29^{+0.24}_{-0.22}$ & $-13.70^{+0.05}_{-0.04}$ \\ \\
            DM + ScatteringGP ($\rm log_{10}A = -13.5$) & $0.98^{+0.01}_{-0.01}$ & $0.55^{+0.17}_{-0.18}$ & $-13.55^{+0.06}_{-0.06}$ & $3.78^{+0.02}_{-0.02}$ & $3.35^{+0.24}_{-0.23}$ & $-13.69^{+0.05}_{-0.04}$ \\ \\
            DM + ScatteringGP ($\rm log_{10}A = -12.5$) & $0.98^{+0.01}_{-0.01}$ & $0.61^{+0.17}_{-0.17}$ & $-12.54^{+0.06}_{-0.06}$ & $3.798^{+0.002}_{-0.002}$ & $3.34^{+0.25}_{-0.23}$ & $-13.69^{+0.05}_{-0.04}$ \\ 
            \bottomrule
        \end{tabular}
    \end{center}
    \caption{Recovered parameter values from different simulations. Here, the DM and Centroid signals were obtained from the IISM simulations, whereas the ScatteringGP signal was simulated as a Gaussian process with varying amplitudes.}
    \label{tab:unusual_chromaticity}
\end{table*}

\section{Conclusion}
\label{sec:conclusion}

The achievable precision in pulsar timing experiments is increasing with the use of increasingly sensitive telescopes. Given this improvement, accurate modelling of subdominant noise terms, such as scattering delays, is becoming essential. However, it is still unclear whether current strategies to account for IISM delays are producing reliable measurements. 
Such an ambiguity may pose a challenge to PTA collaborations utilising future highly sensitive telescopes such as SKA and DSA-2000. In this work, we have developed methods to test the efficacy of the current IISM noise modelling techniques employed by PTA collaborations. These methods facilitate preliminary diagnostics to establish whether the measured delay variations originate in the IISM. We developed these methods based on simulations of multipath propagation of radio waves through the turbulent IISM. 

The simulations have also provided deeper insights into the expected characteristics of the scattering delays as captured in the long-term pulsar timing data sets. The results show that in some circumstances, the scattering delays in the pulsar timing data sets can be correlated with the DM delays according to the expression \ref{eq:tref_propto_sqdmdr}. The degree of correlation depends on the nature of anisotropy in the IISM.
This suggests that measurements of DM and scattering variations can be used to probe the structure of the IISM. The spectral analysis of the simulated IISM delays shows that the strength and the spectral steepness of the scattering delays can vary based on the nature of anisotropy; however, the scattering delays are consistently weaker and spectrally shallower than the DM delays. Our simulations have also highlighted that there can be significant decorrelation in the scattering delays across radio frequencies arising from anisotropic scattering conditions. This behaviour can be important for wideband pulsar timing techniques where its effect might be significant.  
For example, the PPTA collaboration in their third data release has detected excess low frequency band noise in some highly scattered pulsars like PSR J1643$-$1224 \citep{2023ApJ...951L...7R}. Although the exact reason for this is not yet understood, it is possible that the band noise may have arisen from decorrelating scattering variations, whose effect became noticeable at lower frequencies. 
 
The findings from the simulations were used as a basis to test the efficacy of the noise modelling techniques, through a detailed study of the IISM delays from the MPTA 4.5-year data set. We found mixed evidence for the robustness of the current IISM noise models. For instance, we found that the measured strength of the scattering variations in the MPTA data set was weaker than the DM variations for most of the pulsars, which is consistent with the outcomes of our simulations. We also found a potential occurrence of scattering events along the lines of sight of PSRs J1431$-$5740 and J1802$-$2124, which were identified through correlations between flux and DM time series. One of the possible explanations for these events could be the presence of plasma underdensities along their lines of sight.

However, we also found evidence that improvements could be made to the noise modelling process. For instance, time-series analysis shows that the scattering delay exhibits relatively slow variations as opposed to the fast variations seen in the simulations. 
We also identified occasional instances of anticorrelation between the DM and scattering variations, which we attribute to an artefact of the fitting process. Apart from that, it was noticed that when noise models included chromatic Gaussian events we could identify strong anticorrelations between DM and scattering variations. Here, we also note the lack of weak but consistent anticorrelations between scattering variations and flux density variations which were found in the simulations. Although the reason behind this unexpected behaviour is not understood, some of it can be attributed to the covariance between the DM and scattering noise models. Both these terms are strongly frequency dependent and show signs of signal exchange between them. The current PTA IISM models cannot distinguish between the two signals efficiently. A similar behaviour has also been reported by \citet{2024A&A...692A.170I} who show that the estimation of DM variations when modelled as DMGP can be influenced by other red noise processes in the data.

Through simulation we have also demonstrated a possible explanation for steep chromaticity observed in the MPTA 4.5-year data set.
The simulation shows that the steep chromaticities measured in the PTA noise analyses could be physical as they arise due to complex interplay between DM and chromatic noise models when the scattering variations are non-Gaussian in nature. However, we also note that the steep chromaticities of the scattering variations may be impacted by other factors, such as the choice of two-dimensional templates to measure the times of arrival. Such measurement can be strongly covariant with other radio frequency dependent parameters such as the so-called FD parameters and DM and its higher order derivatives.

From this study, it can be concluded that further improvements in the scattering noise models are required to accurately measure the scattering delays. The improved model can consider the possibility of frequency decorrelation in the scattering delays, as observed from our simulations. We may need to develop alternate methods, or improve upon the existing framework of Gaussian processes, to take into account the non-Gaussianity of the scattering delays.

\section*{Acknowledgements}

ADK, DJR, and RMS acknowledge support from the ARC Centre of Excellence for Gravitational Wave Discovery (CE170100004 and CE230100016). RMS acknowledges support through ARC Future Fellowship FT190100155. MTM acknowledges support from the NANOGrav Collaboration's National Science Foundation grant. We thank Aurélien Chalumea for their useful comments on this work. The MeerKAT telescope is operated by the South African Radio Astronomy Observatory (SARAO), which is a facility of the National Research Foundation, an agency of the Department of Science and Innovation. PTUSE was developed with support from the Australian SKA Office and Swinburne University of Technology, with financial contributions from the MeerTime collaboration members. This work used the OzSTAR national facility at Swinburne University of Technology. OzSTAR is funded by Swinburne University of Technology and the National Collaborative Research Infrastructure Strategy (NCRIS).

\section*{Data Availability}
All data used in this work is available courtesy of AAO Data Central (\url{https://datacentral.org.au/}) at \url{https://doi.org/10.57891/j0vh-5g31}. The data generated through simulations will be made available on reasonable request to the corresponding author.

\bsp	
\label{lastpage}
\end{document}